\documentclass[aps,pra,twocolumn,showpacs,superscriptaddress,longbibliography]{revtex4-1}
\usepackage{graphicx}    
\usepackage{dcolumn}    
\usepackage{bm}             
\usepackage{amssymb}   
\usepackage{amsmath}    
\usepackage{subfigure}    
\usepackage{braket}
\usepackage{bm}
\usepackage{mathrsfs}
\usepackage{dsfont}
\usepackage[export]{adjustbox}
\usepackage[toc,page]{appendix}
\begin{document}

\title{Universal hybrid quantum computing in trapped ions}

\author{R. T. Sutherland}
\email{robert.sutherland@utsa.edu}
\affiliation{Department of Electrical and Computer Engineering, Department of Physics and Astronomy, University of Texas at San Antonio, San Antonio, TX 78249, USA}
\author{R. Srinivas}
\affiliation{Department  of  Physics,  University  of  Oxford,  Clarendon  Laboratory, Parks  Road,  Oxford  OX1  3PU,  U.K.}

\date{\today}

\begin{abstract}
Using discrete and continuous variable subsystems, hybrid approaches to quantum information could enable more quantum computational power for the same physical resources. Here, we propose a hybrid scheme that can be used to generate the necessary Gaussian and non-Gaussian operations for universal continuous variable quantum computing in trapped ions. This scheme utilizes two \textit{linear} spin-motion interactions to generate a broad set of \textit{non-linear} effective spin-motion interactions including one and two mode squeezing, beam splitter, and trisqueezing operations in trapped ion systems. We discuss possible experimental implementations using laser-based and laser-free approaches.
\end{abstract}
\maketitle

\section{Introduction}
Quantum information is typically encoded in a discrete set of eigenvalues, such as two-level qubit systems. Practical computation using such systems will require error correction, which may need up to millions of physical qubits, potentially limiting near-term applications \cite{fowler_2012}. Continuous variable quantum computation (CVQC) \cite{lloyd_1999, gottesman_2001, braunstein_2005, lau_2016} offers an alternative approach by storing information in states with a continuum of eigenvalues, enabling higher density encoding for the same physical resources. This approach, however, requires Gaussian operations (displacement, squeezing, etc) and at least one non-Gaussian operation (trisqueezing) \cite{lloyd_1999}. Experimentally, it has thus far been challenging to implement a scheme with the flexibility to perform both types of operations unilaterally. This challenge could be addressed with hybrid platforms that incorporate both discrete and continuous variables, utilizing the advantages of each system \cite{andersen_2015,lau_2016}.

Trapped ions\textemdash a leading platform for quantum information \cite{cirac_1995, monroe_1995,nielsen_2010, haffner_2008,blatt_2008,harty_2014,ballance_2016,gaebler_2016, srinivas_2021}, quantum simulation \cite{porras_2004, jurcevic_2017,sutherland_2019_2}, and quantum metrology \cite{heinzen_1990, wineland_1994, burd_2019, brewer_2019, mccormick_2019} \textemdash are a prime candidate for CVQC \cite{lau_2012} and hybrid quantum computing \cite{andersen_2015,lau_2016,fluhmann_2019}. The Jaynes-Cummings-type interaction between the (discrete) internal states of ions coupled to their (continuous) motional states can generate Gaussian \cite{meekhof_1996,zeng_2002, kienzler_2015} and non-Gaussian \cite{leibfried_2002} operations. The strength of this spin-motion coupling is typically determined by the Lamb-Dicke factor $\eta \equiv \Delta k x_{0}$, where $\Delta k$ is the effective wave number of the applied field and $x_{0}$ is the extent of the ground state motional wave packet; typically, $\eta$ is $\sim$\,0.1 for laser-based interactions and $\sim\, 0.0001$ for laser-free interactions. Existing schemes that generate non-linear interactions with respect to their motional coupling are $\propto \eta^2$ or higher-order, making them significantly weaker than $\propto \eta$ interactions. These higher-order interactions include the state dependent beam splitter and one or two mode squeezing operations, and can also be used to generate non-Gaussian states \cite{gan_2020,drechsler_2020,cardoso_2021}. Here, we introduce a new scheme that scales linearly with respect to $\eta$, enabling stronger, higher-order interactions, and can also be used to implement multiple types of operations using the same control fields, reducing experimental overhead. We show that we can use two, simultaneously applied, spin-dependent displacement interactions \cite{molmer_1999,molmer_2000,leibfried_2003}, typically used in geometric phase gates, to selectively generate effective couplings that are non-linear in their motional component; we will thus refer to this as a geometric phase gate interaction. 

This paper is organized as follows: In Sec.~\ref{sec:continuous}, we show that one can selectively implement a broad set of effective Hamiltonians, representing Gaussian and non-Gaussian interactions, by simultaneously applying two geometric phase gate interactions to one ion, where the spin-components of the interactions do not commute. We can select which effective Hamiltonian to implement by simply adjusting the frequencies of the phase-gate interactions. In Sec.~\ref{sec:microwave}, we present a method for implementing our scheme in a laser-free system with a single spin-motion coupling interaction. In Sec.~\ref{sec:results}, we compare numerical simulations of the (exact) geometric phase gate Hamiltonians with the (approximate) effective Hamiltonians, which converge for long enough gate durations. In Sec.~\ref{sec:conclusion}, we give conclusions and prospects for future work.

\section{Theory}

In this section, we describe how two non-commuting phase-gate interactions can create Gaussian and non-Gaussian operations, sufficient for universal CVQC \cite{lloyd_1999,supplemental}. Such phase-gate interactions are typically used in trapped-ion experiments to generate entanglement with or without lasers. We also show that this scheme can be implemented in a laser-free experiment with a single radiofrequency gradient \cite{sutherland_2019,srinivas_2018,sutherland_2019_2}. We consider a single spin, with states $\ket{\downarrow}$ and $\ket{\uparrow}$, whose interaction is represented by Pauli operators $\hat{\sigma}_{\alpha}$, where $\alpha\in\{x, y, z\}$. The spin is coupled to one or two harmonic oscillator modes $j$, whose interactions are represented with an annihilation(creation) operator $\hat{a}_{j}(\hat{a}_{j}^{\dagger})$.

\subsection{Spin-dependent non-linear spin-motion coupling}\label{sec:continuous}
In this section, we discuss applying interactions that are linear in $\hat{a}_{j}$, to generate effective spin-motion interactions that are non-linear in $\hat{a}_{j}$. We analyze our effective Hamiltonians using the Magnus expansion for a unitary time-propagator acting under Schr\"{o}dinger's equation \cite{magnus_1954}:
\begin{eqnarray}\label{eq:magnus}
\hat{U}(t) &\simeq & \exp\Big(\frac{-i}{\hbar}\int^{t}_{0}dt_{1}\hat{H}_{1} - \frac{1}{2\hbar^{2}}\int^{t}_{0}\int^{t_{1}}_{0}dt_{1}dt_{2}\big[\hat{H}_{1},\hat{H}_{2}\big] \nonumber \\
&+& \frac{i}{6\hbar^{3}}\int^{t}_{0}\int^{t_{1}}_{0}\int^{t_{2}}_{0}dt_{1}dt_{2}dt_{3}\Big\{\Big[\hat{H}_{1},\big[\hat{H}_{2},\hat{H}_{3}\big]\Big] \nonumber \\
&+& \Big[\hat{H}_{3},\big[\hat{H}_{2},\hat{H}_{1}\big]\Big]\Big\}\Big),
\end{eqnarray}
where $\hat{H}_{k}\equiv \hat{H}(t_{k})$ represents the Hamiltonian describing the system at time $t_{k}$. We have truncated the expansion after the $3^{rd}$-order, leaving analysis of higher-order interactions to future work. In the following, we will discuss effective Hamiltonians $\hat{H}_{eff}$, where $\hat{U}(t)\simeq \exp({-it\hat{H}_{eff}/\hbar})$.

We begin with a Hamiltonian that represents two geometric phase gate interactions acting on a single ion in the interaction picture with respect to the ion's qubit and motional energies:
\begin{eqnarray}\label{eq:gate_ham}
\hat{H}(t) = \hbar\Omega_{\alpha}\hat{\sigma}_{\alpha}\hat{a}_{j}e^{-i\Delta t} + \hbar\Omega_{\alpha^{\prime}}\hat{\sigma}_{\alpha^{\prime}}\hat{a}_{j^{\prime}}e^{-i(n\Delta t + \phi)} + c.c., \nonumber \\
\end{eqnarray}
where $\Omega_{\alpha,\alpha^{\prime}}$ are the Rabi frequencies of the corresponding spin-motion coupling terms, $\Delta = 2\pi/t_{i}$ is the detuning, $n$ is an integer, and $\phi$ is an arbitrary phase. Note that $\Omega_{\alpha,\alpha^{\prime}}$ both scale linearly with the Lamb-Dicke factor $\eta$. We assume $\alpha\ne\alpha^{\prime}$. The first-order term in the Magnus expansion is 0 for integer multiples of $t_i$. The total interaction duration is $t_{f} = K t_{i}$, where $K$ is an integer.

Plugging Eq.~(\ref{eq:gate_ham}) into Eq.~(\ref{eq:magnus}), setting $n=\pm 1$, and solving for a duration $t_{f}$ gives:
\begin{eqnarray}\label{eq:second_order}
\hat{U}_{2}(t_{f}) &=& \exp\Big(\Omega_{2}t_{f}\varepsilon_{\alpha\alpha^{\prime}\alpha^{\prime\prime}}\hat{\sigma}_{\alpha^{\prime\prime}}\Big\{\delta_{n,-1}(\hat{a}^{\dagger}_{j}\hat{a}^{\dagger}_{j^{\prime}}e^{i\phi} - \hat{a}_{j}\hat{a}_{j^{\prime}}e^{-i\phi})  \nonumber \\
&&~~~~~~+ \delta_{n,1}(\hat{a}^{\dagger}_{j}\hat{a}_{j^{\prime}}e^{-i\phi} - 
\hat{a}_{j}\hat{a}^{\dagger}_{j^{\prime}}e^{i\phi}) \nonumber \\
&&~~~~~~+ \delta_{j,j^{\prime}}\delta_{n,1}\cos(\phi)\Big\}\Big) + \mathcal{O}([\Omega_{\alpha}/\Delta]^{3}),
\end{eqnarray}

\noindent where $\Omega_{2}\equiv 2\Omega_{\alpha}\Omega_{\alpha^{\prime}}/\Delta$. For the error term, we assume $\Omega_{\alpha}$ and $\Omega_{\alpha^{\prime}}$ are the same order of magnitude. Here, the $2^{\text{nd}}$-order terms in the Magnus expansion are the largest contributions to the overall dynamics of the system. By appropriate choice of $n$, $\phi$, $j$, and $j^{\prime}$ we can create a set of interactions $\hat{H}_{eff}$ that are spin-dependent and $2^{\text{nd}}$-order in $\hat{a}_{j},\hat{a}_{j^{\prime}}$ (see Table \ref{table:eff_hams}), including one- (Sec.~\ref{sec:one_squeeze}) and two-mode (Sec.~\ref{sec:two_squeeze}) squeezing, as well as beam splitter (Sec.~\ref{sec:beam_split}) interactions.

Similarly, for $n=\pm 2$, the $3^{\text{rd}}$-order terms in the Magnus expansion are the largest contribution to the dynamics; the $1^{\text{st}}$- and $2^{\text{nd}}$-order terms integrate to zero. Plugging Eq.~(\ref{eq:gate_ham}) into Eq.~(\ref{eq:magnus}) again, we get:
\begin{eqnarray}\label{eq:third_order}
\hat{U}_{3}(t_{f}) &=& \exp\Big(-i\Omega_{3}t_{f}\hat{\sigma}_{\alpha^{\prime}}\Big\{\delta_{n,2}(\hat{a}_{j}\hat{a}^{\dagger}_{j^{\prime}}\hat{a}_{j}e^{i\phi} + \hat{a}^{\dagger}_{j}\hat{a}_{j^{\prime}}\hat{a}^{\dagger}_{j}e^{-i\phi}) \nonumber \\
&&~~~~~~~+ \delta_{n,-2}(\hat{a}_{j}^{\dagger 2}\hat{a}^{\dagger}_{j^{\prime}}e^{i\phi} + \hat{a}^{2}_{j}\hat{a}_{j^{\prime}}e^{-i\phi}) \Big\} \Big) \nonumber \\
&&~~~~~~~+ \mathcal{O}([\Omega_{\alpha}/\Delta]^{4}), \nonumber \\
\end{eqnarray}
where $\Omega_{3}\equiv 2\Omega_{\alpha^{\prime}}\Omega_{\alpha}^{2}/\Delta^{2}$. This time-evolution operator corresponds to  spin-dependent effective Hamiltonians that are $3^{\text{rd}}$-order in $\hat{a}_{j},\hat{a}_{j^{\prime}}$ (see Table.~\ref{table:eff_hams}), which includes trisqueezing (Sec.~\ref{sec:trisqueezing}).

\begin{table*}[bt]
\begin{center}
\begin{tabular}{|l| c | c | c|}
 \hline
 \textbf{Effective Hamiltonian} & 
 \textbf{Gaussian} &
 \textbf{n} & $\mathbf{j}=\textbf{j}^{\mathbf{\prime}}$ \\ [0.5ex] 
 \hline\hline
 
 $\hat{H}_{eff} \simeq i\hbar\Omega_{2}\varepsilon_{\alpha\alpha^{\prime}\alpha^{\prime\prime}}\hat{\sigma}_{\alpha^{\prime\prime}}\Big(\hat{a}_{j}^{\dagger 2}e^{i\phi} - \hat{a}^{2}_{j}e^{-i\phi} \Big)$ & yes & -1 & yes \\ 
 
 \hline
 $\hat{H}_{eff} \simeq i\hbar\Omega_{2}\varepsilon_{\alpha\alpha^{\prime}\alpha^{\prime\prime}}\hat{\sigma}_{\alpha^{\prime\prime}}\Big(\hat{a}^{\dagger}_{j}\hat{a}^{\dagger}_{j^{\prime}}e^{i\phi} - \hat{a}_{j}\hat{a}_{j^{\prime}}e^{-i\phi} \Big)$ & yes & -1 & no \\
 
 \hline
 $\hat{H}_{eff} \simeq i\hbar\Omega_{2}\varepsilon_{\alpha\alpha^{\prime}\alpha^{\prime\prime}}\hat{\sigma}_{\alpha^{\prime\prime}}\Big(\hat{a}_{j}^{\dagger}\hat{a}_{j}e^{-i\phi} - \hat{a}_{j}\hat{a}^{\dagger}_{j}e^{i\phi} +\cos{\phi}\Big)$ & yes & 1 & yes \\
 
 \hline
 $\hat{H}_{eff} \simeq i\hbar\Omega_{2}\varepsilon_{\alpha\alpha^{\prime}\alpha^{\prime\prime}}\hat{\sigma}_{\alpha^{\prime\prime}}\Big(\hat{a}_{j}^{\dagger}\hat{a}_{j^{\prime}}e^{-i\phi} - \hat{a}_{j}\hat{a}^{\dagger}_{j^{\prime}}e^{i\phi}\Big)$ & yes & 1 & no \\
 
 \hline
  $\hat{H}_{eff} \simeq \hbar\Omega_{3}\hat{\sigma}_{\alpha^{\prime}}\Big(\hat{a}_{j}^{\dagger 3}e^{i\phi} + \hat{a}^{ 3}_{j}e^{-i\phi}\Big)$ & no & -2 & yes \\
  
 \hline
 $\hat{H}_{eff} \simeq \hbar\Omega_{3}\hat{\sigma}_{\alpha^{\prime}}\Big(\hat{a}_{j}^{\dagger 2}\hat{a}_{j^{\prime}}^{\dagger}e^{i\phi} + \hat{a}^{2}_{j}\hat{a}_{j^{\prime}}e^{-i\phi}\Big)$ & no & -2 & no \\
 
 \hline

$\hat{H}_{eff} \simeq \hbar\Omega_{3}\hat{\sigma}_{\alpha^{\prime}}\Big(\hat{a}_{j}\hat{a}_{j}^{\dagger}\hat{a}_{j}e^{i\phi} + \hat{a}_{j}^{\dagger}\hat{a}_{j}\hat{a}^{\dagger}_{j}e^{-i\phi}\Big)$ & no & 2 & yes \\

 \hline
 $\hat{H}_{eff} \simeq \hbar\Omega_{3}\hat{\sigma}_{\alpha^{\prime}}\Big(\hat{a}_{j}^{2}\hat{a}_{j^{\prime}}^{\dagger}e^{i\phi} + \hat{a}_{j}^{2 \dagger}\hat{a}_{j^{\prime}}e^{-i\phi}\Big)$ & no & 2 & no \\
 \hline
\end{tabular}
\caption{List of effective Hamiltonians $\hat{H}_{eff}$ that can be generated by choosing the values of $n$, $\phi$, $j$, and $j^{\prime}$ in Eq.~(\ref{eq:gate_ham}). Setting $n = \pm 1$ generates forms of $\hat{H}_{eff}$ that correspond to the $2^{\text{nd}}$-order term in the Magnus expansion, producing Gaussian operations on $1$ or $2$ phonon modes, with a Rabi frequency $\Omega_{2}\equiv 2\Omega_{\alpha}\Omega_{\alpha^{\prime}}/\Delta$. Setting $n=\pm 2$ generates forms of $\hat{H}_{eff}$ that correspond to the $3^{\text{rd}}$-order terms in the Magnus expansion, producing non-Gaussian operations on $1$ or $2$ modes, with Rabi frequency $\Omega_{3}\equiv 2\Omega_{\alpha^{\prime}}\Omega_{\alpha}^{2}/\Delta^{2}$.}\label{table:eff_hams}
\end{center}
\end{table*}

In both Eq.~(\ref{eq:second_order}) and Eq.~(\ref{eq:third_order}), the time evolution primarily corresponds to a selected $\hat{H}_{eff}$. The undesired, higher-order terms in the Magnus expansion decrease faster with $\Delta$ than $\hat{H}_{eff}$. Therefore, the time evolution converges to that of $\hat{H}_{eff}$ as the iterations $K$ of the interaction and $\Delta$ increase, such that $\Omega_{2}t_{f}$ or $\Omega_{3}t_{f}$ is fixed. Thus, $\hat{U}(t_f)\rightarrow\exp({-\frac{i}{\hbar} \hat{H}_{eff}t_f})$ for large values of $K$. As $t_{f}=Kt_{i}=2\pi K/\Delta$, for $2^{\text{nd}}$-order interactions this results in
\begin{equation}
    \frac{4\pi\Omega_{\alpha}\Omega_{\alpha^{\prime}}K}{\Delta^{2}} = \Omega_{2}t_{f},
\end{equation}
which must be held constant. Thus, as $K$ increases from $1$, $\Delta\rightarrow \Delta K^{1/2}$ and $t_{f}\rightarrow t_{f} K^{1/2}$. Similarly, for $3^{\text{rd}}$-order interactions, we see that
\begin{equation}
    \frac{4\pi \Omega^{2}_{\alpha^{\prime}}\Omega_{\alpha}K}{\Delta^{3}} = \Omega_{3}t_{f},
\end{equation}
must be held constant. Therefore, increasing the value of $K$ from $1$ gives $\Delta\rightarrow \Delta K^{1/3}$ and $t_{f}\rightarrow t_{f}K^{2/3}$. Therefore, for large values of $K$, the evolution of the interactions reported in Table~\ref{table:eff_hams} can be made accurate to an arbitrary degree. In Sec.~\ref{sec:results}, we verify this behaviour through direct numerical integration of Eq.~(\ref{eq:gate_ham}). 

Since the convergence criteria for both the $2^{\text{nd}}$- and $3^{\text{rd}}$-order interactions depend only on $\Omega_{\alpha}/\Delta$, and $\Omega_{\alpha}$ is linear with respect to $\eta$, $\Delta$ is also linear with respect to $\eta$. Thus, $\Omega_{2}$ and $\Omega_{3}$ are linear with respect to $\eta$ as well; this makes our scheme, to the best of our knowledge, the first method for generating spin-motion coupling that is non-linear in $\hat{a}_{j}$ with field interactions that are only linear in $\eta$. This linearity is crucial for laser-free based approaches, where $\eta$ is small and non-linear interactions are hard to generate.

\subsubsection*{Physical Interpretation}

In the above discussion, a clear pattern emerged: the largest non-zero contribution to the overall dynamics of the system comes from terms in the Magnus expansion that oscillate at opposite frequencies to one another. When $n=1$ in Eq.~(\ref{eq:gate_ham}), for example, the $\hat{a}_{j,j^{\prime}}$ and $\hat{a}^{\dagger}_{j,j^{\prime}}$ terms oscillate with frequencies $-\Delta$ and $\Delta$, respectively, leading to $\hat{H}_{eff}\propto \hat{a}^{\dagger}_{j^{\prime}}\hat{a}_{j} + c.c.$. If $n=-1$, the opposite is true, leading to $\hat{H}_{eff}\propto \hat{a}_{j^{\prime}}\hat{a}_{j} + c.c.$. We also see the same behaviour for $3^{\text{rd}}$-order terms in the Magnus expansion when $n=\pm 2$. If we consider that the values of $\Delta$ and $n\Delta$ represent the detuning of the gate fields from their respective transitions, we can interpret this apparent pattern as a result of the \textit{conservation of energy}. This means that $\hat{H}_{eff}$ describes the leading-order energy conserving transition in the Magnus expansion. From this perspective, the physics we describe here parallels that of a Raman transition, where the spin-dependent displacement acts as an auxiliary state. In Sec.~\ref{sec:results} we show that, similar to a Raman transition, the larger the detuning from the auxiliary state, the more the time evolution corresponds to $\hat{H}_{eff}$, due to the reduced effect off-resonant processes. This leads to the question of whether a more general pattern exists, where, for example, larger values of $n$ in Eq.~(\ref{eq:gate_ham}) generate forms of $\hat{H}_{eff}$ that are higher-order in $\hat{a}_{j,j^{\prime}}$ beyond the $3^{\text{rd}}$-order interactions shown here. In this work, we only postulate that such a pattern exists, leaving a rigorous proof to future work. 

\subsection{Laser-free implementation with radiofrequency gradient}\label{sec:microwave}

Equation~(\ref{eq:gate_ham}) may be implemented with multiple pairs of lasers, symmetrically detuned around the transition frequency of the qubit. The same interaction may also be implemented without lasers, using multiple pairs of gradients oscillating close to the qubit frequency~\cite{ospelkaus_2008,ospelkaus_2011}. In this subsection, we present a method of implementing Eq.~(\ref{eq:gate_ham}) in a laser-free system using a single radiofrequency gradient in addition to a pair of weak, symmetrically detuned microwave fields \cite{sutherland_2019,srinivas_2018,srinivas_2021}. Such a scheme would simplify the experimental, overhead as multiple gradients are typically hard to generate.

In the interaction picture with respect to the motion and qubit frequencies, the Hamiltonian for such a microwave-driven system, after the rotating wave approximation, takes the form \cite{sutherland_2019}:
\begin{eqnarray}\label{eq:laser_free_lab}
\hat{H}_{r}(t) &=& 2\hbar\Omega_{\mu}\hat{\sigma}_{x}\cos(\delta t) \nonumber \\
&&+ 2\hat{\sigma}_{z}\cos(\omega_{g}t)\sum_{j}\hbar\Omega_{g,j}\Big\{\hat{a}_{j}e^{-i\omega_{j}t} + \hat{a}_{j}^{\dagger}e^{i\omega_{j}t}\Big\}, \nonumber \\
\end{eqnarray}
where $\Omega_{g,j}$ is the Rabi frequency of the radiofrequency gradient coupled to one or two motional modes $j$, $\Omega_{\mu}$ is the Rabi frequency of the symmetrically detuned microwave pair, and $\delta$ is the magnitude of the detuning of each microwave pair from the qubit frequency. We have assumed the microwave pair is polarized in the $x$ direction. 

We analyze the dynamics of Eq.~(\ref{eq:laser_free_lab}) by transforming into the interaction picture with respect to the bichromatic microwave pair \cite{roos_2008,jonathan_2000, sutherland_2019}:
\begin{eqnarray}
\hat{H}_{I}(t) = \hat{U}_{I}^{\dagger}(t)\hat{H}_{r}(t)\hat{U}_{I}(t) + i\hbar\dot{\hat{U}}_{I}^{\dagger}(t)\hat{U}_{I}(t),
\end{eqnarray}
where the frame transformation $\hat{U}_{I}(t)$ is given by:
\begin{eqnarray}
\hat{U}_{I}(t) = \exp\Big\{-\frac{2i\Omega_{\mu}\sin(\delta t)}{\delta}\hat{\sigma}_{x} \Big\}.
\end{eqnarray}
Here, $\hat{H}_{I}(t)$ takes the form:
\begin{eqnarray}\label{eq:interaction_laser_free}
\hat{H}_{I}(t) &=& 2\hbar \cos(\omega_{g}t)\Big\{\sum_{j}\Omega_{g,j}(\hat{a}_{j} e^{-i\omega_{j}t} + \hat{a}_{j}^{\dagger}e^{i\omega_{j}t}) \Big\}\times \nonumber \\
&&\Big\{ \hat{\sigma}_{z}\Big[J_{0}\Big(\frac{4\Omega_{\mu}}{\delta}\Big)+~2\sum_{m=1}^{\infty}J_{2m}\Big( \frac{4\Omega_{\mu}}{\delta}\Big)\cos(2m\delta t) \Big] \nonumber \\
&& +~2\hat{\sigma}_{y}\sum^{\infty}_{m=1}J_{2m-1}\Big(\frac{4\Omega_{\mu}}{\delta}\Big)\sin([2m-1]\delta t) \Big\},
\end{eqnarray}
where $J_{m}$ represents the $m^{\text{th}}$ Bessel function of the $1^{\text{st}}$ kind, which we assume has an argument of $4\Omega_{\mu}/\delta$ from here on. As $\delta \gg \Omega_{g,j}$ in typical laser-free experiments, Eq.~(\ref{eq:interaction_laser_free}) shows that there is an infinite set of potential spin-motion interactions that can be generated by tuning an integer multiple of $\delta$ to be near any of the $\omega_{j}\pm\omega_{g}$ sidebands. Choosing $\delta$ such that it is an even multiple of $\omega_{j}\pm\omega_{g}$ produces a $\hat{\sigma}_{z}$ coupling to the spin, while an odd multiple of $\delta$ produces a $\hat{\sigma}_{y}$ coupling \cite{sutherland_2019,sutherland_2020}. Finally, we must ensure that the dynamics one expects from $\hat{H}_{I}(t)$ correspond to that of $\hat{H}_{r}(t)$. If the bichromatic microwave pair is ramped on and off slowly compared to $1/\delta$, $\hat{U}_{I}(t_{f})\rightarrow \hat{I}$ and the time-propagator that results from $\hat{H}_{r}(t)$ converges to that of $\hat{H}_{I}(t)$. We can, therefore, we can consider $\hat{H}_{I}(t)$ exclusively when discussing the dynamics of $\hat{H}_{r}(t)$ \cite{sutherland_2019,sutherland_2020}.

Reference~\cite{sutherland_2020} discusses methods to set the values of $\delta$ and $\omega_{g}$ to generate an entangling interaction that is insensitive to decoherence of the qubit and motion. We implement a similar idea here to produce a Hamiltonian that takes the form of Eq.~(\ref{eq:gate_ham}). Assuming that the values of $\omega_{j}$ and $\omega_{j^{\prime}}$ are fixed, we generate Eq.~(\ref{eq:gate_ham}) by setting even and odd multiples of $\delta$ to be near the $\omega_{j}-\omega_{g}$ and $\omega_{j^{\prime}} + \omega_{g}$ sidebands, respectively. As an example, we choose $\delta$ and $\omega_g$ to satisfy:
\begin{eqnarray}\label{eq:two_un}
2\delta &=& (\omega_{j} - \omega_{g}) - \Delta, \nonumber \\
3\delta &=& (\omega_{j^{\prime}} + \omega_{g}) - n\Delta,
\end{eqnarray}
where in the above equation $j^{\prime}$ may or may not refer to the same mode as $j$. Upon making the rotating wave approximation, this gives: 
\begin{eqnarray}
\hat{H}_{I}(t) \simeq \hbar\Big\{\Omega_{g,j}J_{2}\hat{\sigma}_{z}\hat{a}_{j}e^{-i\Delta t} - i\Omega_{g,j'}J_{3}\hat{\sigma}_{y}\hat{a}_{j^{\prime}}e^{-in\Delta t}\Big\} + c.c., \nonumber \\
\end{eqnarray}
taking the form of Eq.~(\ref{eq:gate_ham}) wherein $\Omega_{\alpha}=\Omega_{g,j}J_{2}$, $\Omega_{\alpha^{\prime}}=\Omega_{g,j'}J_{3}$, and $\phi=\pi/2$. Thus, our scheme enables non-linear spin-motion coupling, without lasers, using a single gradient. 

\section{Numerical simulations}\label{sec:results}
Here, we show that, for a selection of the $\hat{H}_{eff}$ shown in Table~\ref{table:eff_hams}, the time-dynamics produced by $\hat{H}_{eff}$ converge to those of Eq.~(\ref{eq:gate_ham}) when $\Delta$ is large compared to $\Omega_{\alpha,\alpha^{\prime}}$. Here, we assume that $\Omega_{\alpha}$ and $\Omega_{\alpha^{\prime}}$ have the same value $\Omega$. We also assume that $\alpha = y$ and $\alpha^{\prime} = x$. We show the convergence to $\hat{H}_{eff}$ in Figs.~\ref{fig:one_mode_sq}-\ref{fig:trisqueeze} with the `fidelity' $\mathcal{F}$ of each operation, defined as $\mathcal{F}\equiv|\braket{T|\psi(t)}|^{2}$, where $\ket{T}$ is the target state obtained through direct integration of $\hat{H}_{eff}$, and $\ket{\psi(t)}$ is the state obtained through direct integration of Eq.~(\ref{eq:gate_ham}). After the simulation, we can approximate experimental run-times for both laser-based and laser-free experiments, using Rabi frequencies of $\Omega/2\pi = 10 ~\mathrm{kHz}$ and $\Omega/2\pi = 1~ \mathrm{kHz}$, respectively, which are within the parameter regime of current experiments \cite{ballance_2016,gaebler_2016, srinivas_2018}. 

\subsection{One-mode squeezing}\label{sec:one_squeeze}
\begin{figure}
\includegraphics[width=0.5\textwidth]{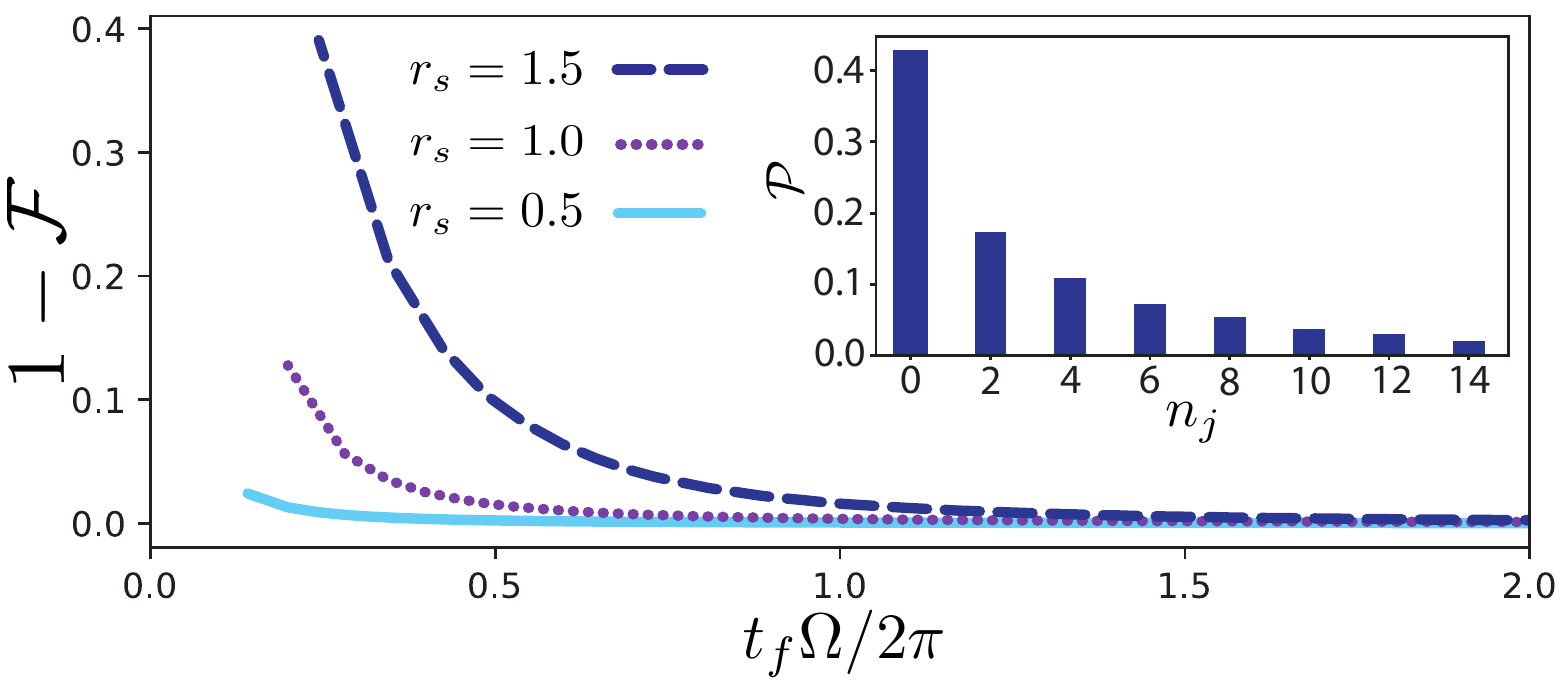}
\centering
\caption{One-mode squeezing operation generated by Eq.~(\ref{eq:gate_ham}). We plot the infidelity $1-\mathcal{F}$ versus normalized gate duration $t_{f}\Omega$, while keeping $\Omega_{2}t_{f}$ constant. This dependence is shown for squeezing parameters $2\Omega_{2}t_{f}\equiv r_{s}$ of 1.5 (blue dashed), 1.0 (purple dotted), and $0.5$ (solid light blue). For large values of $t_{f}$, $1-\mathcal{F}\rightarrow 0$, as described in the text. The inset shows the probability $\mathcal{P}$ of phonon state $\ket{n_j}$ for a squeezed state with $r_{s} = 1.5$, generated by integrating Eq.~(\ref{eq:gate_ham}) for $t_f\Omega/2\pi\simeq 2.2$.}
\label{fig:one_mode_sq}
\end{figure}

One mode squeezing has been shown to enhance quantum metrology~\cite{caves_1981,heinzen_1990,aasi_2013,burd_2019}, the speed of entangling gates~\cite{ge_2019, burd_2020}, and is an integral component of CVQC~\cite{lloyd_1999}. In trapped ions, squeezing of the motion can be generated through modulation of the trapping potential at twice the motional frequency \cite{heinzen_1990,burd_2019}, a diabatic change of the trapping frequency \cite{heinzen_1990,wittemer_2019}, a pair of laser beams with a difference frequency that is twice the motional frequency~\cite{meekhof_1996, dupays_2021}, or reservoir engineering~\cite{kienzler_2015}. Here, we present an alternative scheme, where each geometric phase gate interaction in Eq.~(\ref{eq:gate_ham}) operates on the same mode ($j=j^{\prime}$) with opposite detunings $\Delta$ from one another ($n=-1$). This interaction produces a spin-dependent squeezing operation:
\begin{eqnarray}\label{eq:one_mode_squeeze}
\hat{H}_{eff} \simeq i\hbar\Omega_{2}\hat{\sigma}_{z}\Big(\hat{a}_{j}^{\dagger 2}e^{i\phi} - \hat{a}^{2}_{j}e^{-i\phi} \Big).
\end{eqnarray}
 This spin-dependent squeezing operation can be combined with single qubit rotations and spin-measurements to create non-Gaussian states \cite{drechsler_2020}.

In Fig.~\ref{fig:one_mode_sq}, we numerically integrate both Eq.~(\ref{eq:gate_ham}) and Eq.~(\ref{eq:one_mode_squeeze}), showing $1-\mathcal{F}$ versus $t_{f}$ for squeezing parameters $r_{s}\equiv 2\Omega_{2}t_{f}$ of $0.5$, $1.0$, and $1.5$. For all three calculations, the results of Eq.~(\ref{eq:gate_ham}) and Eq.~(\ref{eq:one_mode_squeeze}) converge to one another, showing that our scheme can produce an effective squeezing operation to arbitrary accuracy for large enough values of $t_{f}$. We can use the results shown in Fig.~\ref{fig:one_mode_sq} to approximate the experimental run-time needed to induce a given value of $r_{s}$ and $\mathcal{F}$. For example, a squeezing parameter of $r_{s}=1.5$ with fidelity $\mathcal{F}\simeq 0.99$ has a normalized gate duration of $t_{f}\Omega/2\pi = 1.2$. This corresponds to 120 $\mu$s for laser-based ($\Omega/2\pi = 10~\mathrm{kHz}$), and $1.2\,$ms for laser-free ($\Omega/2\pi=1~\mathrm{kHz}$) implementations.

\subsection{Two-mode squeezing}\label{sec:two_squeeze}

\begin{figure}
\includegraphics[width=0.5\textwidth]{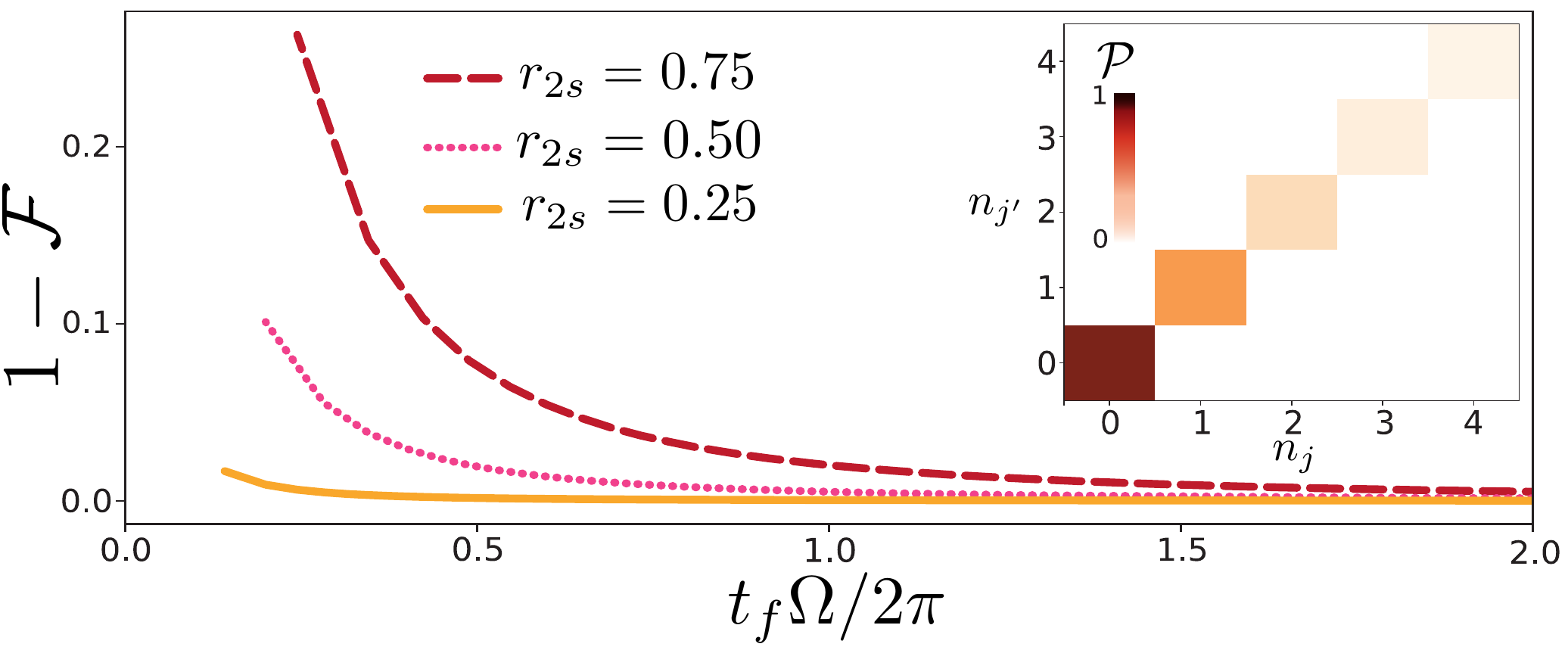}
\centering
\caption{Two-mode squeezing operation generated by Eq.~(\ref{eq:gate_ham}). Here we show the infidelity $1-\mathcal{F}$ versus the normalized gate duration $t_{f}\Omega/2\pi$. This dependence is shown for squeezing parameters $r_{2s}$ of 0.75 (red dashed), 0.5 (pink dotted), and 0.25 (orange solid). For larger values of $t_{f}$, all three calculations converge to that of $\hat{H}_{eff}$. The inset shows the probabilities $\mathcal{P}$ of occupying a state with $n_{j}$ phonons in the $j$ mode and $n_{j^{\prime}}$ phonons in the $j^{\prime}$ mode for $r_{2s}=0.75$ and $t_{f}\Omega/2\pi\simeq 1.7$.}
\label{fig:two_mode_sq}
\end{figure}

Two-mode squeezing generates an entangled state between two bosonic modes \cite{agarwal_2012}. It has been used as a resource for CVQC, quantum cryptography, as well as quantum teleportation \cite{braunstein_2005,weedbrook_2012,grosshans_2003,furusawa_1998}, and is potentially useful for quantum metrology \cite{cardoso_2021}. Experiments have been demonstrated with optical \cite{furusawa_1998,grosshans_2003} and microwave \cite{eichler_2011} photons, but not the phonon modes of a trapped ion crystal. Two-mode squeezing in trapped ions has been proposed through the non-linear motional coupling of the ions to laser fields oscillating at $\omega_{j}+\omega_{j^{\prime}}$ \cite{zeng_2002, cardoso_2021}; our approach offers an alternative that uses only linear motional coupling. Implementing Eq.~(\ref{eq:gate_ham}), where each geometric phase gate interaction operates on different modes ($j\neq j^{\prime}$) and each of the interactions have equal and opposite detunings ($n=-1$), the effective interaction is:
\begin{eqnarray}\label{eq:two_mode_squeeze}
\hat{H}_{eff} \simeq i\hbar\Omega_{2}\hat{\sigma}_{z}\Big(\hat{a}^{\dagger}_{j}\hat{a}^{\dagger}_{j^{\prime}}e^{i\phi} - \hat{a}_{j}\hat{a}_{j^{\prime}}e^{-i\phi} \Big),
\end{eqnarray}
representing spin-dependent two-mode squeezing. Since this Hamiltonian is spin-dependent, it may be combined with single-qubit operations and observations of the spin to create the superpositions of multiple two-mode squeezed states described in Ref.~\cite{cardoso_2021}. 

In Fig.~\ref{fig:two_mode_sq}, we numerically integrate Eq.~(\ref{eq:gate_ham}) and Eq.~(\ref{eq:two_mode_squeeze}) for $n=-1$ and $j\neq j^{\prime}$, showing $1-\mathcal{F}$ versus the normalized gate duration $t_{f}\Omega/2\pi$ for squeezing parameters $r_{2s}\equiv \Omega_{2}t_{f}$ \cite{agarwal_2012} of $0.25$, $0.5$, and $0.75$; here, there is a factor of two difference in the definitions of $r_{2s}$ and $r_{s}$ \cite{agarwal_2012}. For all three calculations, the results of Eq.~(\ref{eq:gate_ham}) and Eq.~(\ref{eq:two_mode_squeeze}) converge, showing that our scheme is capable of producing an effective two-mode squeezing operation to arbitrary accuracy for large enough values of $t_{f}$. For $r_{2s}=0.75$ and $\mathcal{F} \simeq 0.99$, $t_{f}\Omega/2\pi = 1.4$, giving experimental run-times of $t_{f}\simeq 140~ \mu\mathrm{s}$ and 1.4$\,$ms for laser-based ($\Omega/2\pi = 10~\mathrm{kHz}$) and laser-free parameters ($\Omega/2\pi = 1~\mathrm{kHz}$), respectively.

\subsection{Beam splitter}\label{sec:beam_split}

\begin{figure}
\includegraphics[width=0.5\textwidth]{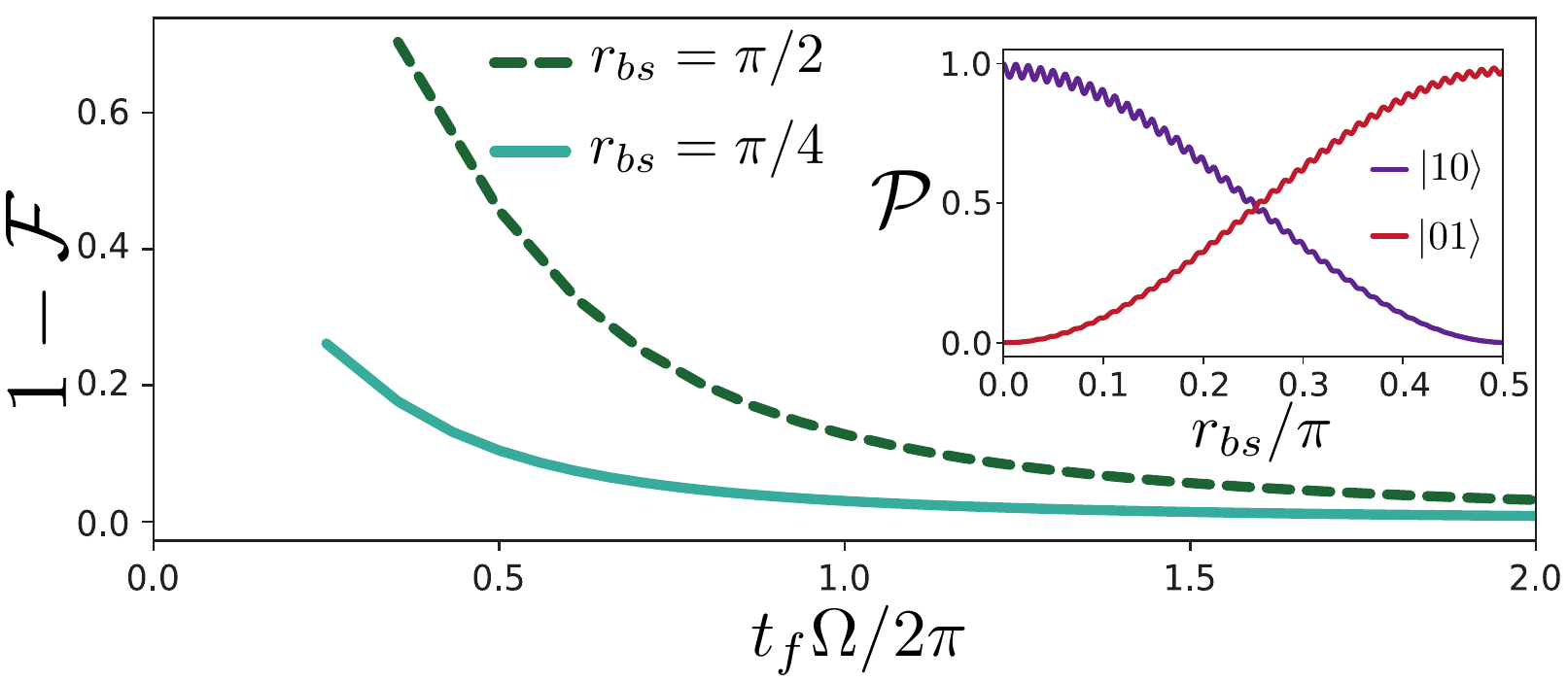}
\centering
\caption{Beam-splitter operation generated by Eq.~(\ref{eq:gate_ham}). Here, we show the infidelity $1-\mathcal{F}$ versus normalized gate duration $t_{f}\Omega/2\pi$. This dependence is shown for two squeezing parameters $r_{bs} = \pi/4$ (solid turquoise) and $r_{bs}=\pi/2$ (dashed green), for a system initialized to the state $\ket{\downarrow}\ket{n_j=1, n_{j'}=0}$. The inset shows the probabilities of the phonon subspace being in the state $\ket{10}$ and $\ket{01}$, versus time, for a $r_{bs}=\pi/2$ interaction such that $t_{f}\Omega/2\pi \simeq 2.5$.}
\label{fig:beam_splitter}
\end{figure}

The beam splitter interaction swaps the states of two boson modes \cite{agarwal_2012}. This interaction is useful for CVQC \cite{lloyd_1999,braunstein_2005,weedbrook_2012}, and can also be used to cool multiple motional modes of trapped ion crystals \cite{gorman_2014,lau_2014}. This interaction has been experimentally demonstrated in trapped ions by modulating the voltages of the trap electrodes \cite{gorman_2014}, and with a running optical lattice \cite{gan_2020}. Instead, we here implement Eq.~(\ref{eq:gate_ham}), where each geometric phase gate interaction operates on different modes ($j\neq j^{\prime}$) and each interaction has the same detuning ($n=1$). The resulting interaction is
\begin{equation}\label{eq:beam_split}
    \hat{H}_{eff} \simeq i\hbar\Omega_{2}\hat{\sigma}_{z}\Big(\hat{a}_{j}\hat{a}^{\dagger}_{j^{\prime}}e^{i\phi}- \hat{a}_{j}^{\dagger}\hat{a}_{j^{\prime}}e^{-i\phi}\Big),
\end{equation}
representing a spin-dependent beam splitter.

In Fig.~\ref{fig:beam_splitter}, we numerically integrate Eq.~(\ref{eq:gate_ham}), for $n=1$ and $j\neq j^{\prime}$, and Eq.~(\ref{eq:beam_split}), showing $1-\mathcal{F}$ versus $t_{f}\Omega/2\pi$. The figure compares two calculations where the system is initialized to the state $\ket{\downarrow}\ket{n_j=1,n_{j'}=0}$, for ${r_{bs}\equiv \Omega_{2}t_{f}}$ of $\pi/4$ and $\pi/2$. The inset of the figure shows the probabilities of finding the motion in the states $\ket{10}$ and $\ket{01}$, traced over the spin degree-of-freedom, for $t_{f}\Omega/2\pi\simeq 2.5$ and $r_{bs}=\pi/2$. For a value of $r_{bs}=\pi/4$, a value of $t_{f}\Omega/2\pi\simeq 1.8$ is needed to achieve $\mathcal{F}\simeq 0.99$. This duration is $t_{f} \simeq 180~\mu\mathrm{s}$ for laser-based parameters ($\Omega/2\pi = 10~\mathrm{kHz}$), and  $t_{f}\simeq 1.8~\mathrm{ms}$ for laser-free parameters ($\Omega/2\pi\simeq 1~\mathrm{kHz}$).

\subsection{Trisqueezing}\label{sec:trisqueezing}

\begin{figure}
\includegraphics[width=0.5\textwidth]{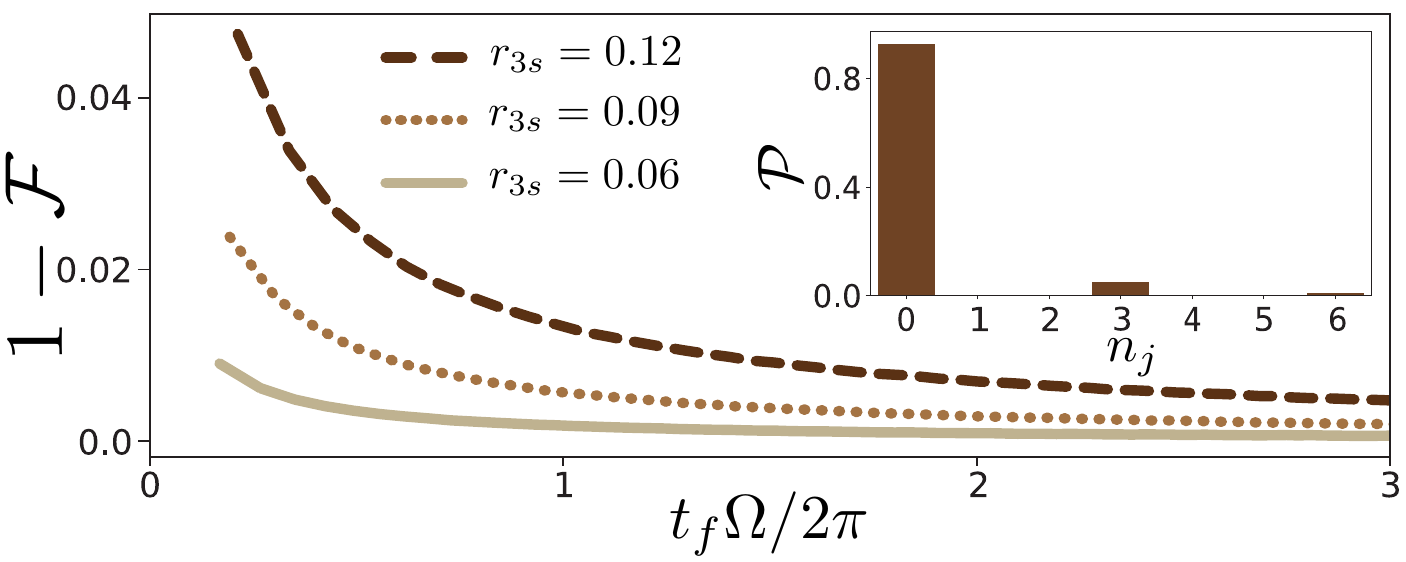}
\centering
\caption{Trisqueezing operation generated by Eq.~(\ref{eq:gate_ham}) . We show the infidelity $1 -\mathcal{F}$ versus normalized gate duration $t_{f}\Omega/2\pi$ for three squeezing parameters $r_{3s} = 0.12$ (brown dashed), $r_{3s}=0.09$ (light brown dotted), and $r_{3s}=0.06$ (tan solid). The inset shows the probabilities $\mathcal{P}$ of each phonon state $n_j$ for $r_{3s}=0.12$ and $t_{f}\Omega/2\pi\simeq 2.9$ generated by Eq.~(\ref{eq:gate_ham}), showing that only states with phonon numbers that are multiples of $3$ are populated.}
\label{fig:trisqueeze}
\end{figure}

Universal CVQC requires at least one non-Gaussian operation \cite{lloyd_1999}. For $n=-2$ and $j=j^{\prime}$ in Eq.~(\ref{eq:gate_ham}), we generate the effective interaction:
\begin{eqnarray}\label{eq:trisqueeze}
    \hat{H}_{eff} \simeq \hbar\Omega_{3}\hat{\sigma}_{x}\Big(\hat{a}_{j}^{\dagger 3}e^{i\phi} + \hat{a}^{ 3}_{j}e^{i\phi}\Big), 
\end{eqnarray}
which is the trisqueezing Hamiltonian \cite{fisher_1984, hillery_1984,braunstein_1987,hillery_1990}, which has recently been demonstrated using microwave photons~\cite{chang_2020}. Here, we stay in a regime where the squeezing parameter is small $t_f\Omega_{3}\equiv r_{3s} \leq 0.12$. Importantly, implementing the non-Gaussian trisqueezing interaction using our scheme has no additional experimental overhead compared to the Gaussian interactions discussed above.

In Fig.~\ref{fig:trisqueeze}, we numerically integrate Eq.~(\ref{eq:gate_ham}) starting in the motional ground state, and compare the results to Eq.~(\ref{eq:trisqueeze}) by showing $1-\mathcal{F}$ versus the normalized time $t_{f}\Omega/2\pi$. The inset of the figure shows the probability of the motional mode having a given number of phonons $n_j$, tracing over the spin degree-of-freedom for an implementation of Eq.~(\ref{eq:gate_ham}) with and $t_{f}\Omega/2\pi\simeq 2.9$; this shows that, as expected, only states with phonon numbers that are multiples of three have non-zero populations. For $r_{3s}=0.12$, a gate duration of $t_{f}\Omega/2\pi\simeq 1.4$ is needed to achieve $\mathcal{F}\simeq 0.99$ which corresponds to $t_{f} \simeq 140~\mu\mathrm{s}$ and $t_{f}\simeq 1.4~\mathrm{ms}$ using laser-based ($\Omega/2\pi = 10~\mathrm{kHz}$), and laser-free parameters ($\Omega/2\pi = 1~\mathrm{kHz}$), respectively. \\

\section{Conclusion}\label{sec:conclusion}
In conclusion, we have introduced a new hybrid scheme for implementing spin-motion interactions that are \textit{non-linear} in their motional component, using only spin-motion interactions that are \textit{linear} in their motional component. Through the application two geometric phase gate interactions acting on a single ion, we showed that it is possible to create a broad set of Gaussian and non-Gaussian operations, sufficient for universal quantum computation over continuous variables. The type of operation can be chosen by simply changing the frequencies of the individual phase gate interactions. We also proposed an implementation of our scheme in a laser-free setup using a single radiofrequency gradient. Finally, we verified the validity of our scheme numerically, and provided approximate experimental run-times. While most experimental demonstrations of non-linear coupling to a boson mode have only focused on one type of interaction, all the interactions discussed here can be implemented in the same experimental setup.

\section*{Acknowledgements}
We would like to thank D. T. C. Allcock, C. J.~Ballance, O.~Bazavan, M. Minder, S. Saner, and D. J. Wineland for thoughtful discussions.

\bibliographystyle{apsrev4-1}
\bibliography{biblio}

\begin{thebibliography}{65}%
\makeatletter
\providecommand \@ifxundefined [1]{%
 \@ifx{#1\undefined}
}%
\providecommand \@ifnum [1]{%
 \ifnum #1\expandafter \@firstoftwo
 \else \expandafter \@secondoftwo
 \fi
}%
\providecommand \@ifx [1]{%
 \ifx #1\expandafter \@firstoftwo
 \else \expandafter \@secondoftwo
 \fi
}%
\providecommand \natexlab [1]{#1}%
\providecommand \enquote  [1]{``#1''}%
\providecommand \bibnamefont  [1]{#1}%
\providecommand \bibfnamefont [1]{#1}%
\providecommand \citenamefont [1]{#1}%
\providecommand \href@noop [0]{\@secondoftwo}%
\providecommand \href [0]{\begingroup \@sanitize@url \@href}%
\providecommand \@href[1]{\@@startlink{#1}\@@href}%
\providecommand \@@href[1]{\endgroup#1\@@endlink}%
\providecommand \@sanitize@url [0]{\catcode `\\12\catcode `\$12\catcode
  `\&12\catcode `\#12\catcode `\^12\catcode `\_12\catcode `\%12\relax}%
\providecommand \@@startlink[1]{}%
\providecommand \@@endlink[0]{}%
\providecommand \url  [0]{\begingroup\@sanitize@url \@url }%
\providecommand \@url [1]{\endgroup\@href {#1}{\urlprefix }}%
\providecommand \urlprefix  [0]{URL }%
\providecommand \Eprint [0]{\href }%
\providecommand \doibase [0]{http://dx.doi.org/}%
\providecommand \selectlanguage [0]{\@gobble}%
\providecommand \bibinfo  [0]{\@secondoftwo}%
\providecommand \bibfield  [0]{\@secondoftwo}%
\providecommand \translation [1]{[#1]}%
\providecommand \BibitemOpen [0]{}%
\providecommand \bibitemStop [0]{}%
\providecommand \bibitemNoStop [0]{.\EOS\space}%
\providecommand \EOS [0]{\spacefactor3000\relax}%
\providecommand \BibitemShut  [1]{\csname bibitem#1\endcsname}%
\let\auto@bib@innerbib\@empty
\bibitem [{\citenamefont {Fowler}\ \emph {et~al.}(2012)\citenamefont {Fowler},
  \citenamefont {Mariantoni}, \citenamefont {Martinis},\ and\ \citenamefont
  {Cleland}}]{fowler_2012}%
  \BibitemOpen
  \bibfield  {author} {\bibinfo {author} {\bibfnamefont {A.~G.}\ \bibnamefont
  {Fowler}}, \bibinfo {author} {\bibfnamefont {M.}~\bibnamefont {Mariantoni}},
  \bibinfo {author} {\bibfnamefont {J.~M.}\ \bibnamefont {Martinis}}, \ and\
  \bibinfo {author} {\bibfnamefont {A.~N.}\ \bibnamefont {Cleland}},\ }\href
  {\doibase 10.1103/PhysRevA.86.032324} {\bibfield  {journal} {\bibinfo
  {journal} {Phys. Rev. A}\ }\textbf {\bibinfo {volume} {86}},\ \bibinfo
  {pages} {032324} (\bibinfo {year} {2012})}\BibitemShut {NoStop}%
\bibitem [{\citenamefont {Lloyd}\ and\ \citenamefont
  {Braunstein}(1999)}]{lloyd_1999}%
  \BibitemOpen
  \bibfield  {author} {\bibinfo {author} {\bibfnamefont {S.}~\bibnamefont
  {Lloyd}}\ and\ \bibinfo {author} {\bibfnamefont {S.~L.}\ \bibnamefont
  {Braunstein}},\ }\href {\doibase 10.1103/PhysRevLett.82.1784} {\bibfield
  {journal} {\bibinfo  {journal} {Phys. Rev. Lett.}\ }\textbf {\bibinfo
  {volume} {82}},\ \bibinfo {pages} {1784} (\bibinfo {year}
  {1999})}\BibitemShut {NoStop}%
\bibitem [{\citenamefont {Gottesman}\ \emph {et~al.}(2001)\citenamefont
  {Gottesman}, \citenamefont {Kitaev},\ and\ \citenamefont
  {Preskill}}]{gottesman_2001}%
  \BibitemOpen
  \bibfield  {author} {\bibinfo {author} {\bibfnamefont {D.}~\bibnamefont
  {Gottesman}}, \bibinfo {author} {\bibfnamefont {A.}~\bibnamefont {Kitaev}}, \
  and\ \bibinfo {author} {\bibfnamefont {J.}~\bibnamefont {Preskill}},\ }\href
  {\doibase 10.1103/PhysRevA.64.012310} {\bibfield  {journal} {\bibinfo
  {journal} {Phys. Rev. A}\ }\textbf {\bibinfo {volume} {64}},\ \bibinfo
  {pages} {012310} (\bibinfo {year} {2001})}\BibitemShut {NoStop}%
\bibitem [{\citenamefont {Braunstein}\ and\ \citenamefont
  {Van~Loock}(2005)}]{braunstein_2005}%
  \BibitemOpen
  \bibfield  {author} {\bibinfo {author} {\bibfnamefont {S.~L.}\ \bibnamefont
  {Braunstein}}\ and\ \bibinfo {author} {\bibfnamefont {P.}~\bibnamefont
  {Van~Loock}},\ }\href@noop {} {\bibfield  {journal} {\bibinfo  {journal}
  {Rev. of Mod. Phys.}\ }\textbf {\bibinfo {volume} {77}},\ \bibinfo {pages}
  {513} (\bibinfo {year} {2005})}\BibitemShut {NoStop}%
\bibitem [{\citenamefont {Lau}\ and\ \citenamefont {Plenio}(2016)}]{lau_2016}%
  \BibitemOpen
  \bibfield  {author} {\bibinfo {author} {\bibfnamefont {H.-K.}\ \bibnamefont
  {Lau}}\ and\ \bibinfo {author} {\bibfnamefont {M.~B.}\ \bibnamefont
  {Plenio}},\ }\href@noop {} {\bibfield  {journal} {\bibinfo  {journal} {Phys.
  Rev. Lett.}\ }\textbf {\bibinfo {volume} {117}},\ \bibinfo {pages} {100501}
  (\bibinfo {year} {2016})}\BibitemShut {NoStop}%
\bibitem [{\citenamefont {Andersen}\ \emph {et~al.}(2015)\citenamefont
  {Andersen}, \citenamefont {Neergaard-Nielsen}, \citenamefont {Van~Loock},\
  and\ \citenamefont {Furusawa}}]{andersen_2015}%
  \BibitemOpen
  \bibfield  {author} {\bibinfo {author} {\bibfnamefont {U.~L.}\ \bibnamefont
  {Andersen}}, \bibinfo {author} {\bibfnamefont {J.~S.}\ \bibnamefont
  {Neergaard-Nielsen}}, \bibinfo {author} {\bibfnamefont {P.}~\bibnamefont
  {Van~Loock}}, \ and\ \bibinfo {author} {\bibfnamefont {A.}~\bibnamefont
  {Furusawa}},\ }\href@noop {} {\bibfield  {journal} {\bibinfo  {journal} {Nat.
  Phys.}\ }\textbf {\bibinfo {volume} {11}},\ \bibinfo {pages} {713} (\bibinfo
  {year} {2015})}\BibitemShut {NoStop}%
\bibitem [{\citenamefont {Cirac}\ and\ \citenamefont
  {Zoller}(1995)}]{cirac_1995}%
  \BibitemOpen
  \bibfield  {author} {\bibinfo {author} {\bibfnamefont {J.~I.}\ \bibnamefont
  {Cirac}}\ and\ \bibinfo {author} {\bibfnamefont {P.}~\bibnamefont {Zoller}},\
  }\href {\doibase 10.1103/PhysRevLett.74.4091} {\bibfield  {journal} {\bibinfo
   {journal} {Phys. Rev. Lett.}\ }\textbf {\bibinfo {volume} {74}},\ \bibinfo
  {pages} {4091} (\bibinfo {year} {1995})}\BibitemShut {NoStop}%
\bibitem [{\citenamefont {Monroe}\ \emph {et~al.}(1995)\citenamefont {Monroe},
  \citenamefont {Meekhof}, \citenamefont {King}, \citenamefont {Itano},\ and\
  \citenamefont {Wineland}}]{monroe_1995}%
  \BibitemOpen
  \bibfield  {author} {\bibinfo {author} {\bibfnamefont {C.}~\bibnamefont
  {Monroe}}, \bibinfo {author} {\bibfnamefont {D.~M.}\ \bibnamefont {Meekhof}},
  \bibinfo {author} {\bibfnamefont {B.~E.}\ \bibnamefont {King}}, \bibinfo
  {author} {\bibfnamefont {W.~M.}\ \bibnamefont {Itano}}, \ and\ \bibinfo
  {author} {\bibfnamefont {D.~J.}\ \bibnamefont {Wineland}},\ }\href {\doibase
  10.1103/PhysRevLett.75.4714} {\bibfield  {journal} {\bibinfo  {journal}
  {Phys. Rev. Lett.}\ }\textbf {\bibinfo {volume} {75}},\ \bibinfo {pages}
  {4714} (\bibinfo {year} {1995})}\BibitemShut {NoStop}%
\bibitem [{\citenamefont {Nielsen}\ and\ \citenamefont
  {Chuang}(2010)}]{nielsen_2010}%
  \BibitemOpen
  \bibfield  {author} {\bibinfo {author} {\bibfnamefont {M.~A.}\ \bibnamefont
  {Nielsen}}\ and\ \bibinfo {author} {\bibfnamefont {I.~L.}\ \bibnamefont
  {Chuang}},\ }\href@noop {} {\emph {\bibinfo {title} {Quantum computation and
  quantum information}}}\ (\bibinfo  {publisher} {Cambridge University Press},\
  \bibinfo {year} {2010})\BibitemShut {NoStop}%
\bibitem [{\citenamefont {H{\"a}ffner}\ \emph {et~al.}(2008)\citenamefont
  {H{\"a}ffner}, \citenamefont {Roos},\ and\ \citenamefont
  {Blatt}}]{haffner_2008}%
  \BibitemOpen
  \bibfield  {author} {\bibinfo {author} {\bibfnamefont {H.}~\bibnamefont
  {H{\"a}ffner}}, \bibinfo {author} {\bibfnamefont {C.~F.}\ \bibnamefont
  {Roos}}, \ and\ \bibinfo {author} {\bibfnamefont {R.}~\bibnamefont {Blatt}},\
  }\href@noop {} {\bibfield  {journal} {\bibinfo  {journal} {Phys. Rep.}\
  }\textbf {\bibinfo {volume} {469}},\ \bibinfo {pages} {155} (\bibinfo {year}
  {2008})}\BibitemShut {NoStop}%
\bibitem [{\citenamefont {Blatt}\ and\ \citenamefont
  {Wineland}(2008)}]{blatt_2008}%
  \BibitemOpen
  \bibfield  {author} {\bibinfo {author} {\bibfnamefont {R.}~\bibnamefont
  {Blatt}}\ and\ \bibinfo {author} {\bibfnamefont {D.~J.}\ \bibnamefont
  {Wineland}},\ }\href@noop {} {\bibfield  {journal} {\bibinfo  {journal}
  {Nature}\ }\textbf {\bibinfo {volume} {453}},\ \bibinfo {pages} {1008}
  (\bibinfo {year} {2008})}\BibitemShut {NoStop}%
\bibitem [{\citenamefont {Harty}\ \emph {et~al.}(2014)\citenamefont {Harty},
  \citenamefont {Allcock}, \citenamefont {Ballance}, \citenamefont {Guidoni},
  \citenamefont {Janacek}, \citenamefont {Linke}, \citenamefont {Stacey},\ and\
  \citenamefont {Lucas}}]{harty_2014}%
  \BibitemOpen
  \bibfield  {author} {\bibinfo {author} {\bibfnamefont {T.~P.}\ \bibnamefont
  {Harty}}, \bibinfo {author} {\bibfnamefont {D.~T.~C.}\ \bibnamefont
  {Allcock}}, \bibinfo {author} {\bibfnamefont {C.~J.}\ \bibnamefont
  {Ballance}}, \bibinfo {author} {\bibfnamefont {L.}~\bibnamefont {Guidoni}},
  \bibinfo {author} {\bibfnamefont {H.~A.}\ \bibnamefont {Janacek}}, \bibinfo
  {author} {\bibfnamefont {N.~M.}\ \bibnamefont {Linke}}, \bibinfo {author}
  {\bibfnamefont {D.~N.}\ \bibnamefont {Stacey}}, \ and\ \bibinfo {author}
  {\bibfnamefont {D.~M.}\ \bibnamefont {Lucas}},\ }\href {\doibase
  10.1103/PhysRevLett.113.220501} {\bibfield  {journal} {\bibinfo  {journal}
  {Phys. Rev. Lett.}\ }\textbf {\bibinfo {volume} {113}},\ \bibinfo {pages}
  {220501} (\bibinfo {year} {2014})}\BibitemShut {NoStop}%
\bibitem [{\citenamefont {Ballance}\ \emph {et~al.}(2016)\citenamefont
  {Ballance}, \citenamefont {Harty}, \citenamefont {Linke}, \citenamefont
  {Sepiol},\ and\ \citenamefont {Lucas}}]{ballance_2016}%
  \BibitemOpen
  \bibfield  {author} {\bibinfo {author} {\bibfnamefont {C.~J.}\ \bibnamefont
  {Ballance}}, \bibinfo {author} {\bibfnamefont {T.~P.}\ \bibnamefont {Harty}},
  \bibinfo {author} {\bibfnamefont {N.~M.}\ \bibnamefont {Linke}}, \bibinfo
  {author} {\bibfnamefont {M.~A.}\ \bibnamefont {Sepiol}}, \ and\ \bibinfo
  {author} {\bibfnamefont {D.~M.}\ \bibnamefont {Lucas}},\ }\href {\doibase
  10.1103/PhysRevLett.117.060504} {\bibfield  {journal} {\bibinfo  {journal}
  {Phys. Rev. Lett.}\ }\textbf {\bibinfo {volume} {117}},\ \bibinfo {pages}
  {060504} (\bibinfo {year} {2016})}\BibitemShut {NoStop}%
\bibitem [{\citenamefont {Gaebler}\ \emph {et~al.}(2016)\citenamefont
  {Gaebler}, \citenamefont {Tan}, \citenamefont {Lin}, \citenamefont {Wan},
  \citenamefont {Bowler}, \citenamefont {Keith}, \citenamefont {Glancy},
  \citenamefont {Coakley}, \citenamefont {Knill}, \citenamefont {Leibfried},\
  and\ \citenamefont {Wineland}}]{gaebler_2016}%
  \BibitemOpen
  \bibfield  {author} {\bibinfo {author} {\bibfnamefont {J.~P.}\ \bibnamefont
  {Gaebler}}, \bibinfo {author} {\bibfnamefont {T.~R.}\ \bibnamefont {Tan}},
  \bibinfo {author} {\bibfnamefont {Y.}~\bibnamefont {Lin}}, \bibinfo {author}
  {\bibfnamefont {Y.}~\bibnamefont {Wan}}, \bibinfo {author} {\bibfnamefont
  {R.}~\bibnamefont {Bowler}}, \bibinfo {author} {\bibfnamefont {A.~C.}\
  \bibnamefont {Keith}}, \bibinfo {author} {\bibfnamefont {S.}~\bibnamefont
  {Glancy}}, \bibinfo {author} {\bibfnamefont {K.}~\bibnamefont {Coakley}},
  \bibinfo {author} {\bibfnamefont {E.}~\bibnamefont {Knill}}, \bibinfo
  {author} {\bibfnamefont {D.}~\bibnamefont {Leibfried}}, \ and\ \bibinfo
  {author} {\bibfnamefont {D.~J.}\ \bibnamefont {Wineland}},\ }\href {\doibase
  10.1103/PhysRevLett.117.060505} {\bibfield  {journal} {\bibinfo  {journal}
  {Phys. Rev. Lett.}\ }\textbf {\bibinfo {volume} {117}},\ \bibinfo {pages}
  {060505} (\bibinfo {year} {2016})}\BibitemShut {NoStop}%
\bibitem [{\citenamefont {Srinivas}\ \emph {et~al.}(2021)\citenamefont
  {Srinivas}, \citenamefont {Burd}, \citenamefont {Knaack}, \citenamefont
  {Sutherland}, \citenamefont {Kwiatkowski}, \citenamefont {Glancy},
  \citenamefont {Knill}, \citenamefont {Wineland}, \citenamefont {Leibfried},
  \citenamefont {Wilson}, \citenamefont {Allcock},\ and\ \citenamefont
  {Slichter}}]{srinivas_2021}%
  \BibitemOpen
  \bibfield  {author} {\bibinfo {author} {\bibfnamefont {R.}~\bibnamefont
  {Srinivas}}, \bibinfo {author} {\bibfnamefont {S.~C.}\ \bibnamefont {Burd}},
  \bibinfo {author} {\bibfnamefont {H.~M.}\ \bibnamefont {Knaack}}, \bibinfo
  {author} {\bibfnamefont {R.~T.}\ \bibnamefont {Sutherland}}, \bibinfo
  {author} {\bibfnamefont {A.}~\bibnamefont {Kwiatkowski}}, \bibinfo {author}
  {\bibfnamefont {S.}~\bibnamefont {Glancy}}, \bibinfo {author} {\bibfnamefont
  {E.}~\bibnamefont {Knill}}, \bibinfo {author} {\bibfnamefont {D.~J.}\
  \bibnamefont {Wineland}}, \bibinfo {author} {\bibfnamefont {D.}~\bibnamefont
  {Leibfried}}, \bibinfo {author} {\bibfnamefont {A.~C.}\ \bibnamefont
  {Wilson}}, \bibinfo {author} {\bibfnamefont {D.~T.~C.}\ \bibnamefont
  {Allcock}}, \ and\ \bibinfo {author} {\bibfnamefont {D.~H.}\ \bibnamefont
  {Slichter}},\ }\href@noop {} {\bibfield  {journal} {\bibinfo  {journal}
  {arXiv preprint arXiv:2102.12533}\ } (\bibinfo {year} {2021})}\BibitemShut
  {NoStop}%
\bibitem [{\citenamefont {Porras}\ and\ \citenamefont
  {Cirac}(2004)}]{porras_2004}%
  \BibitemOpen
  \bibfield  {author} {\bibinfo {author} {\bibfnamefont {D.}~\bibnamefont
  {Porras}}\ and\ \bibinfo {author} {\bibfnamefont {J.~I.}\ \bibnamefont
  {Cirac}},\ }\href {\doibase 10.1103/PhysRevLett.92.207901} {\bibfield
  {journal} {\bibinfo  {journal} {Phys. Rev. Lett.}\ }\textbf {\bibinfo
  {volume} {92}},\ \bibinfo {pages} {207901} (\bibinfo {year}
  {2004})}\BibitemShut {NoStop}%
\bibitem [{\citenamefont {Jurcevic}\ \emph {et~al.}(2017)\citenamefont
  {Jurcevic}, \citenamefont {Shen}, \citenamefont {Hauke}, \citenamefont
  {Maier}, \citenamefont {Brydges}, \citenamefont {Hempel}, \citenamefont
  {Lanyon}, \citenamefont {Heyl}, \citenamefont {Blatt},\ and\ \citenamefont
  {Roos}}]{jurcevic_2017}%
  \BibitemOpen
  \bibfield  {author} {\bibinfo {author} {\bibfnamefont {P.}~\bibnamefont
  {Jurcevic}}, \bibinfo {author} {\bibfnamefont {H.}~\bibnamefont {Shen}},
  \bibinfo {author} {\bibfnamefont {P.}~\bibnamefont {Hauke}}, \bibinfo
  {author} {\bibfnamefont {C.}~\bibnamefont {Maier}}, \bibinfo {author}
  {\bibfnamefont {T.}~\bibnamefont {Brydges}}, \bibinfo {author} {\bibfnamefont
  {C.}~\bibnamefont {Hempel}}, \bibinfo {author} {\bibfnamefont {B.~P.}\
  \bibnamefont {Lanyon}}, \bibinfo {author} {\bibfnamefont {M.}~\bibnamefont
  {Heyl}}, \bibinfo {author} {\bibfnamefont {R.}~\bibnamefont {Blatt}}, \ and\
  \bibinfo {author} {\bibfnamefont {C.~F.}\ \bibnamefont {Roos}},\ }\href
  {\doibase 10.1103/PhysRevLett.119.080501} {\bibfield  {journal} {\bibinfo
  {journal} {Phys. Rev. Lett.}\ }\textbf {\bibinfo {volume} {119}},\ \bibinfo
  {pages} {080501} (\bibinfo {year} {2017})}\BibitemShut {NoStop}%
\bibitem [{\citenamefont {Sutherland}(2019)}]{sutherland_2019_2}%
  \BibitemOpen
  \bibfield  {author} {\bibinfo {author} {\bibfnamefont {R.~T.}\ \bibnamefont
  {Sutherland}},\ }\href {\doibase 10.1103/PhysRevA.100.061405} {\bibfield
  {journal} {\bibinfo  {journal} {Phys. Rev. A}\ }\textbf {\bibinfo {volume}
  {100}},\ \bibinfo {pages} {061405} (\bibinfo {year} {2019})}\BibitemShut
  {NoStop}%
\bibitem [{\citenamefont {Heinzen}\ and\ \citenamefont
  {Wineland}(1990)}]{heinzen_1990}%
  \BibitemOpen
  \bibfield  {author} {\bibinfo {author} {\bibfnamefont {D.~J.}\ \bibnamefont
  {Heinzen}}\ and\ \bibinfo {author} {\bibfnamefont {D.~J.}\ \bibnamefont
  {Wineland}},\ }\href@noop {} {\bibfield  {journal} {\bibinfo  {journal}
  {Phys. Rev. A}\ }\textbf {\bibinfo {volume} {42}},\ \bibinfo {pages} {2977}
  (\bibinfo {year} {1990})}\BibitemShut {NoStop}%
\bibitem [{\citenamefont {Wineland}\ \emph {et~al.}(1994)\citenamefont
  {Wineland}, \citenamefont {Bollinger}, \citenamefont {Itano},\ and\
  \citenamefont {Heinzen}}]{wineland_1994}%
  \BibitemOpen
  \bibfield  {author} {\bibinfo {author} {\bibfnamefont {D.~J.}\ \bibnamefont
  {Wineland}}, \bibinfo {author} {\bibfnamefont {J.~J.}\ \bibnamefont
  {Bollinger}}, \bibinfo {author} {\bibfnamefont {W.~M.}\ \bibnamefont
  {Itano}}, \ and\ \bibinfo {author} {\bibfnamefont {D.~J.}\ \bibnamefont
  {Heinzen}},\ }\href {\doibase 10.1103/PhysRevA.50.67} {\bibfield  {journal}
  {\bibinfo  {journal} {Phys. Rev. A}\ }\textbf {\bibinfo {volume} {50}},\
  \bibinfo {pages} {67} (\bibinfo {year} {1994})}\BibitemShut {NoStop}%
\bibitem [{\citenamefont {Burd}\ \emph {et~al.}(2019)\citenamefont {Burd},
  \citenamefont {Srinivas}, \citenamefont {Bollinger}, \citenamefont {Wilson},
  \citenamefont {Wineland}, \citenamefont {Leibfried}, \citenamefont
  {Slichter},\ and\ \citenamefont {Allcock}}]{burd_2019}%
  \BibitemOpen
  \bibfield  {author} {\bibinfo {author} {\bibfnamefont {S.~C.}\ \bibnamefont
  {Burd}}, \bibinfo {author} {\bibfnamefont {R.}~\bibnamefont {Srinivas}},
  \bibinfo {author} {\bibfnamefont {J.~J.}\ \bibnamefont {Bollinger}}, \bibinfo
  {author} {\bibfnamefont {A.~C.}\ \bibnamefont {Wilson}}, \bibinfo {author}
  {\bibfnamefont {D.~J.}\ \bibnamefont {Wineland}}, \bibinfo {author}
  {\bibfnamefont {D.}~\bibnamefont {Leibfried}}, \bibinfo {author}
  {\bibfnamefont {D.~H.}\ \bibnamefont {Slichter}}, \ and\ \bibinfo {author}
  {\bibfnamefont {D.~T.~C.}\ \bibnamefont {Allcock}},\ }\href@noop {}
  {\bibfield  {journal} {\bibinfo  {journal} {Science}\ }\textbf {\bibinfo
  {volume} {364}},\ \bibinfo {pages} {1163} (\bibinfo {year}
  {2019})}\BibitemShut {NoStop}%
\bibitem [{\citenamefont {Brewer}\ \emph {et~al.}(2019)\citenamefont {Brewer},
  \citenamefont {Chen}, \citenamefont {Hankin}, \citenamefont {Clements},
  \citenamefont {Chou}, \citenamefont {Wineland}, \citenamefont {Hume},\ and\
  \citenamefont {Leibrandt}}]{brewer_2019}%
  \BibitemOpen
  \bibfield  {author} {\bibinfo {author} {\bibfnamefont {S.~M.}\ \bibnamefont
  {Brewer}}, \bibinfo {author} {\bibfnamefont {J.-S.}\ \bibnamefont {Chen}},
  \bibinfo {author} {\bibfnamefont {A.~M.}\ \bibnamefont {Hankin}}, \bibinfo
  {author} {\bibfnamefont {E.~R.}\ \bibnamefont {Clements}}, \bibinfo {author}
  {\bibfnamefont {C.-W.}\ \bibnamefont {Chou}}, \bibinfo {author}
  {\bibfnamefont {D.~J.}\ \bibnamefont {Wineland}}, \bibinfo {author}
  {\bibfnamefont {D.~B.}\ \bibnamefont {Hume}}, \ and\ \bibinfo {author}
  {\bibfnamefont {D.~R.}\ \bibnamefont {Leibrandt}},\ }\href@noop {} {\bibfield
   {journal} {\bibinfo  {journal} {Phys. Rev. Lett.}\ }\textbf {\bibinfo
  {volume} {123}},\ \bibinfo {pages} {033201} (\bibinfo {year}
  {2019})}\BibitemShut {NoStop}%
\bibitem [{\citenamefont {McCormick}\ \emph {et~al.}(2019)\citenamefont
  {McCormick}, \citenamefont {Keller}, \citenamefont {Burd}, \citenamefont
  {Wineland}, \citenamefont {Wilson},\ and\ \citenamefont
  {Leibfried}}]{mccormick_2019}%
  \BibitemOpen
  \bibfield  {author} {\bibinfo {author} {\bibfnamefont {K.~C.}\ \bibnamefont
  {McCormick}}, \bibinfo {author} {\bibfnamefont {J.}~\bibnamefont {Keller}},
  \bibinfo {author} {\bibfnamefont {S.~C.}\ \bibnamefont {Burd}}, \bibinfo
  {author} {\bibfnamefont {D.~J.}\ \bibnamefont {Wineland}}, \bibinfo {author}
  {\bibfnamefont {A.~C.}\ \bibnamefont {Wilson}}, \ and\ \bibinfo {author}
  {\bibfnamefont {D.}~\bibnamefont {Leibfried}},\ }\href@noop {} {\bibfield
  {journal} {\bibinfo  {journal} {Nature}\ }\textbf {\bibinfo {volume} {572}},\
  \bibinfo {pages} {86} (\bibinfo {year} {2019})}\BibitemShut {NoStop}%
\bibitem [{\citenamefont {Lau}\ and\ \citenamefont {James}(2012)}]{lau_2012}%
  \BibitemOpen
  \bibfield  {author} {\bibinfo {author} {\bibfnamefont {H.-K.}\ \bibnamefont
  {Lau}}\ and\ \bibinfo {author} {\bibfnamefont {D.~F.}\ \bibnamefont
  {James}},\ }\href@noop {} {\bibfield  {journal} {\bibinfo  {journal} {Phys.
  Rev. A}\ }\textbf {\bibinfo {volume} {85}},\ \bibinfo {pages} {062329}
  (\bibinfo {year} {2012})}\BibitemShut {NoStop}%
\bibitem [{\citenamefont {Fl{\"u}hmann}\ \emph {et~al.}(2019)\citenamefont
  {Fl{\"u}hmann}, \citenamefont {Nguyen}, \citenamefont {Marinelli},
  \citenamefont {Negnevitsky}, \citenamefont {Mehta},\ and\ \citenamefont
  {Home}}]{fluhmann_2019}%
  \BibitemOpen
  \bibfield  {author} {\bibinfo {author} {\bibfnamefont {C.}~\bibnamefont
  {Fl{\"u}hmann}}, \bibinfo {author} {\bibfnamefont {T.~L.}\ \bibnamefont
  {Nguyen}}, \bibinfo {author} {\bibfnamefont {M.}~\bibnamefont {Marinelli}},
  \bibinfo {author} {\bibfnamefont {V.}~\bibnamefont {Negnevitsky}}, \bibinfo
  {author} {\bibfnamefont {K.}~\bibnamefont {Mehta}}, \ and\ \bibinfo {author}
  {\bibfnamefont {J.~P.}\ \bibnamefont {Home}},\ }\href@noop {} {\bibfield
  {journal} {\bibinfo  {journal} {Nature}\ }\textbf {\bibinfo {volume} {566}},\
  \bibinfo {pages} {513} (\bibinfo {year} {2019})}\BibitemShut {NoStop}%
\bibitem [{\citenamefont {Meekhof}\ \emph {et~al.}(1996)\citenamefont
  {Meekhof}, \citenamefont {Monroe}, \citenamefont {King}, \citenamefont
  {Itano},\ and\ \citenamefont {Wineland}}]{meekhof_1996}%
  \BibitemOpen
  \bibfield  {author} {\bibinfo {author} {\bibfnamefont {D.~M.}\ \bibnamefont
  {Meekhof}}, \bibinfo {author} {\bibfnamefont {C.}~\bibnamefont {Monroe}},
  \bibinfo {author} {\bibfnamefont {B.~E.}\ \bibnamefont {King}}, \bibinfo
  {author} {\bibfnamefont {W.~M.}\ \bibnamefont {Itano}}, \ and\ \bibinfo
  {author} {\bibfnamefont {D.~J.}\ \bibnamefont {Wineland}},\ }\href@noop {}
  {\bibfield  {journal} {\bibinfo  {journal} {Phys. Rev. Lett.}\ }\textbf
  {\bibinfo {volume} {76}},\ \bibinfo {pages} {1796} (\bibinfo {year}
  {1996})}\BibitemShut {NoStop}%
\bibitem [{\citenamefont {Zeng}\ \emph {et~al.}(2002)\citenamefont {Zeng},
  \citenamefont {Kuang},\ and\ \citenamefont {Gao}}]{zeng_2002}%
  \BibitemOpen
  \bibfield  {author} {\bibinfo {author} {\bibfnamefont {H.-S.}\ \bibnamefont
  {Zeng}}, \bibinfo {author} {\bibfnamefont {L.-M.}\ \bibnamefont {Kuang}}, \
  and\ \bibinfo {author} {\bibfnamefont {K.-L.}\ \bibnamefont {Gao}},\ }\href
  {\doibase https://doi.org/10.1016/S0375-9601(02)00815-0} {\bibfield
  {journal} {\bibinfo  {journal} {Phys. Lett. A}\ }\textbf {\bibinfo {volume}
  {300}},\ \bibinfo {pages} {427} (\bibinfo {year} {2002})}\BibitemShut
  {NoStop}%
\bibitem [{\citenamefont {Kienzler}\ \emph {et~al.}(2015)\citenamefont
  {Kienzler}, \citenamefont {Lo}, \citenamefont {Keitch}, \citenamefont
  {De~Clercq}, \citenamefont {Leupold}, \citenamefont {Lindenfelser},
  \citenamefont {Marinelli}, \citenamefont {Negnevitsky},\ and\ \citenamefont
  {Home}}]{kienzler_2015}%
  \BibitemOpen
  \bibfield  {author} {\bibinfo {author} {\bibfnamefont {D.}~\bibnamefont
  {Kienzler}}, \bibinfo {author} {\bibfnamefont {H.-Y.}\ \bibnamefont {Lo}},
  \bibinfo {author} {\bibfnamefont {B.}~\bibnamefont {Keitch}}, \bibinfo
  {author} {\bibfnamefont {L.}~\bibnamefont {De~Clercq}}, \bibinfo {author}
  {\bibfnamefont {F.}~\bibnamefont {Leupold}}, \bibinfo {author} {\bibfnamefont
  {F.}~\bibnamefont {Lindenfelser}}, \bibinfo {author} {\bibfnamefont
  {M.}~\bibnamefont {Marinelli}}, \bibinfo {author} {\bibfnamefont
  {V.}~\bibnamefont {Negnevitsky}}, \ and\ \bibinfo {author} {\bibfnamefont
  {J.~P.}\ \bibnamefont {Home}},\ }\href@noop {} {\bibfield  {journal}
  {\bibinfo  {journal} {Science}\ }\textbf {\bibinfo {volume} {347}},\ \bibinfo
  {pages} {53} (\bibinfo {year} {2015})}\BibitemShut {NoStop}%
\bibitem [{\citenamefont {Leibfried}\ \emph {et~al.}(2002)\citenamefont
  {Leibfried}, \citenamefont {DeMarco}, \citenamefont {Meyer}, \citenamefont
  {Rowe}, \citenamefont {Ben-Kish}, \citenamefont {Britton}, \citenamefont
  {Itano}, \citenamefont {Jelenkovi{\'c}}, \citenamefont {Langer},
  \citenamefont {Rosenband} \emph {et~al.}}]{leibfried_2002}%
  \BibitemOpen
  \bibfield  {author} {\bibinfo {author} {\bibfnamefont {D.}~\bibnamefont
  {Leibfried}}, \bibinfo {author} {\bibfnamefont {B.}~\bibnamefont {DeMarco}},
  \bibinfo {author} {\bibfnamefont {V.}~\bibnamefont {Meyer}}, \bibinfo
  {author} {\bibfnamefont {M.}~\bibnamefont {Rowe}}, \bibinfo {author}
  {\bibfnamefont {A.}~\bibnamefont {Ben-Kish}}, \bibinfo {author}
  {\bibfnamefont {J.}~\bibnamefont {Britton}}, \bibinfo {author} {\bibfnamefont
  {W.~M.}\ \bibnamefont {Itano}}, \bibinfo {author} {\bibfnamefont
  {B.}~\bibnamefont {Jelenkovi{\'c}}}, \bibinfo {author} {\bibfnamefont
  {C.}~\bibnamefont {Langer}}, \bibinfo {author} {\bibfnamefont
  {T.}~\bibnamefont {Rosenband}},  \emph {et~al.},\ }\href@noop {} {\bibfield
  {journal} {\bibinfo  {journal} {Phys. Rev. Lett.}\ }\textbf {\bibinfo
  {volume} {89}},\ \bibinfo {pages} {247901} (\bibinfo {year}
  {2002})}\BibitemShut {NoStop}%
\bibitem [{\citenamefont {Gan}\ \emph {et~al.}(2020)\citenamefont {Gan},
  \citenamefont {Maslennikov}, \citenamefont {Tseng}, \citenamefont {Nguyen},\
  and\ \citenamefont {Matsukevich}}]{gan_2020}%
  \BibitemOpen
  \bibfield  {author} {\bibinfo {author} {\bibfnamefont {H.~C.~J.}\
  \bibnamefont {Gan}}, \bibinfo {author} {\bibfnamefont {G.}~\bibnamefont
  {Maslennikov}}, \bibinfo {author} {\bibfnamefont {K.-W.}\ \bibnamefont
  {Tseng}}, \bibinfo {author} {\bibfnamefont {C.}~\bibnamefont {Nguyen}}, \
  and\ \bibinfo {author} {\bibfnamefont {D.}~\bibnamefont {Matsukevich}},\
  }\href {\doibase 10.1103/PhysRevLett.124.170502} {\bibfield  {journal}
  {\bibinfo  {journal} {Phys. Rev. Lett.}\ }\textbf {\bibinfo {volume} {124}},\
  \bibinfo {pages} {170502} (\bibinfo {year} {2020})}\BibitemShut {NoStop}%
\bibitem [{\citenamefont {Drechsler}\ \emph {et~al.}(2020)\citenamefont
  {Drechsler}, \citenamefont {Far{\'\i}as}, \citenamefont {Freitas},
  \citenamefont {Schmiegelow},\ and\ \citenamefont {Paz}}]{drechsler_2020}%
  \BibitemOpen
  \bibfield  {author} {\bibinfo {author} {\bibfnamefont {M.}~\bibnamefont
  {Drechsler}}, \bibinfo {author} {\bibfnamefont {M.~B.}\ \bibnamefont
  {Far{\'\i}as}}, \bibinfo {author} {\bibfnamefont {N.}~\bibnamefont
  {Freitas}}, \bibinfo {author} {\bibfnamefont {C.~T.}\ \bibnamefont
  {Schmiegelow}}, \ and\ \bibinfo {author} {\bibfnamefont {J.~P.}\ \bibnamefont
  {Paz}},\ }\href@noop {} {\bibfield  {journal} {\bibinfo  {journal} {Phys.
  Rev. A}\ }\textbf {\bibinfo {volume} {101}},\ \bibinfo {pages} {052331}
  (\bibinfo {year} {2020})}\BibitemShut {NoStop}%
\bibitem [{\citenamefont {Cardoso}\ \emph {et~al.}(2021)\citenamefont
  {Cardoso}, \citenamefont {Rossatto}, \citenamefont {Fernandes}, \citenamefont
  {Higgins},\ and\ \citenamefont {Villas-Boas}}]{cardoso_2021}%
  \BibitemOpen
  \bibfield  {author} {\bibinfo {author} {\bibfnamefont {F.~R.}\ \bibnamefont
  {Cardoso}}, \bibinfo {author} {\bibfnamefont {D.~Z.}\ \bibnamefont
  {Rossatto}}, \bibinfo {author} {\bibfnamefont {G.~P. L.~M.}\ \bibnamefont
  {Fernandes}}, \bibinfo {author} {\bibfnamefont {G.}~\bibnamefont {Higgins}},
  \ and\ \bibinfo {author} {\bibfnamefont {C.~J.}\ \bibnamefont
  {Villas-Boas}},\ }\href@noop {} {\bibfield  {journal} {\bibinfo  {journal}
  {arXiv preprint arXiv:2102.01032}\ } (\bibinfo {year} {2021})}\BibitemShut
  {NoStop}%
\bibitem [{\citenamefont {M\o{}lmer}\ and\ \citenamefont
  {S\o{}rensen}(1999)}]{molmer_1999}%
  \BibitemOpen
  \bibfield  {author} {\bibinfo {author} {\bibfnamefont {K.}~\bibnamefont
  {M\o{}lmer}}\ and\ \bibinfo {author} {\bibfnamefont {A.}~\bibnamefont
  {S\o{}rensen}},\ }\href {\doibase 10.1103/PhysRevLett.82.1835} {\bibfield
  {journal} {\bibinfo  {journal} {Phys. Rev. Lett.}\ }\textbf {\bibinfo
  {volume} {82}},\ \bibinfo {pages} {1835} (\bibinfo {year}
  {1999})}\BibitemShut {NoStop}%
\bibitem [{\citenamefont {S\o{}rensen}\ and\ \citenamefont
  {M\o{}lmer}(2000)}]{molmer_2000}%
  \BibitemOpen
  \bibfield  {author} {\bibinfo {author} {\bibfnamefont {A.}~\bibnamefont
  {S\o{}rensen}}\ and\ \bibinfo {author} {\bibfnamefont {K.}~\bibnamefont
  {M\o{}lmer}},\ }\href {\doibase 10.1103/PhysRevA.62.022311} {\bibfield
  {journal} {\bibinfo  {journal} {Phys. Rev. A}\ }\textbf {\bibinfo {volume}
  {62}},\ \bibinfo {pages} {022311} (\bibinfo {year} {2000})}\BibitemShut
  {NoStop}%
\bibitem [{\citenamefont {Leibfried}\ \emph {et~al.}(2003)\citenamefont
  {Leibfried}, \citenamefont {DeMarco}, \citenamefont {Meyer}, \citenamefont
  {Lucas}, \citenamefont {Barrett}, \citenamefont {Britton}, \citenamefont
  {Itano}, \citenamefont {Jelenkovi{\'c}}, \citenamefont {Langer},
  \citenamefont {Rosenband},\ and\ \citenamefont {Wineland}}]{leibfried_2003}%
  \BibitemOpen
  \bibfield  {author} {\bibinfo {author} {\bibfnamefont {D.}~\bibnamefont
  {Leibfried}}, \bibinfo {author} {\bibfnamefont {B.}~\bibnamefont {DeMarco}},
  \bibinfo {author} {\bibfnamefont {V.}~\bibnamefont {Meyer}}, \bibinfo
  {author} {\bibfnamefont {D.~M.}\ \bibnamefont {Lucas}}, \bibinfo {author}
  {\bibfnamefont {M.}~\bibnamefont {Barrett}}, \bibinfo {author} {\bibfnamefont
  {J.}~\bibnamefont {Britton}}, \bibinfo {author} {\bibfnamefont {W.~M.}\
  \bibnamefont {Itano}}, \bibinfo {author} {\bibfnamefont {B.}~\bibnamefont
  {Jelenkovi{\'c}}}, \bibinfo {author} {\bibfnamefont {C.}~\bibnamefont
  {Langer}}, \bibinfo {author} {\bibfnamefont {T.}~\bibnamefont {Rosenband}}, \
  and\ \bibinfo {author} {\bibfnamefont {D.~J.}\ \bibnamefont {Wineland}},\
  }\href@noop {} {\bibfield  {journal} {\bibinfo  {journal} {Nature}\ }\textbf
  {\bibinfo {volume} {422}},\ \bibinfo {pages} {412} (\bibinfo {year}
  {2003})}\BibitemShut {NoStop}%
\bibitem [{sup()}]{supplemental}%
  \BibitemOpen
  \href@noop {} {}\bibinfo {note} {See supplemental material}\BibitemShut
  {NoStop}%
\bibitem [{\citenamefont {Sutherland}\ \emph {et~al.}(2019)\citenamefont
  {Sutherland}, \citenamefont {Srinivas}, \citenamefont {Burd}, \citenamefont
  {Leibfried}, \citenamefont {Wilson}, \citenamefont {Wineland}, \citenamefont
  {Allcock}, \citenamefont {Slichter},\ and\ \citenamefont
  {Libby}}]{sutherland_2019}%
  \BibitemOpen
  \bibfield  {author} {\bibinfo {author} {\bibfnamefont {R.~T.}\ \bibnamefont
  {Sutherland}}, \bibinfo {author} {\bibfnamefont {R.}~\bibnamefont
  {Srinivas}}, \bibinfo {author} {\bibfnamefont {S.~C.}\ \bibnamefont {Burd}},
  \bibinfo {author} {\bibfnamefont {D.}~\bibnamefont {Leibfried}}, \bibinfo
  {author} {\bibfnamefont {A.~C.}\ \bibnamefont {Wilson}}, \bibinfo {author}
  {\bibfnamefont {D.~J.}\ \bibnamefont {Wineland}}, \bibinfo {author}
  {\bibfnamefont {D.~T.~C.}\ \bibnamefont {Allcock}}, \bibinfo {author}
  {\bibfnamefont {D.~H.}\ \bibnamefont {Slichter}}, \ and\ \bibinfo {author}
  {\bibfnamefont {S.~B.}\ \bibnamefont {Libby}},\ }\href@noop {} {\bibfield
  {journal} {\bibinfo  {journal} {New J. Phys.}\ }\textbf {\bibinfo {volume}
  {21}},\ \bibinfo {pages} {033033} (\bibinfo {year} {2019})}\BibitemShut
  {NoStop}%
\bibitem [{\citenamefont {Srinivas}\ \emph {et~al.}(2019)\citenamefont
  {Srinivas}, \citenamefont {Burd}, \citenamefont {Sutherland}, \citenamefont
  {Wilson}, \citenamefont {Wineland}, \citenamefont {Leibfried}, \citenamefont
  {Allcock},\ and\ \citenamefont {Slichter}}]{srinivas_2018}%
  \BibitemOpen
  \bibfield  {author} {\bibinfo {author} {\bibfnamefont {R.}~\bibnamefont
  {Srinivas}}, \bibinfo {author} {\bibfnamefont {S.~C.}\ \bibnamefont {Burd}},
  \bibinfo {author} {\bibfnamefont {R.~T.}\ \bibnamefont {Sutherland}},
  \bibinfo {author} {\bibfnamefont {A.~C.}\ \bibnamefont {Wilson}}, \bibinfo
  {author} {\bibfnamefont {D.~J.}\ \bibnamefont {Wineland}}, \bibinfo {author}
  {\bibfnamefont {D.}~\bibnamefont {Leibfried}}, \bibinfo {author}
  {\bibfnamefont {D.~T.~C.}\ \bibnamefont {Allcock}}, \ and\ \bibinfo {author}
  {\bibfnamefont {D.~H.}\ \bibnamefont {Slichter}},\ }\href@noop {} {\bibfield
  {journal} {\bibinfo  {journal} {Phys. Rev. Lett.}\ }\textbf {\bibinfo
  {volume} {122}},\ \bibinfo {pages} {163201} (\bibinfo {year}
  {2019})}\BibitemShut {NoStop}%
\bibitem [{\citenamefont {Magnus}(1954)}]{magnus_1954}%
  \BibitemOpen
  \bibfield  {author} {\bibinfo {author} {\bibfnamefont {W.}~\bibnamefont
  {Magnus}},\ }\href@noop {} {\bibfield  {journal} {\bibinfo  {journal}
  {Commun. Pure Appl. Math.}\ }\textbf {\bibinfo {volume} {7}},\ \bibinfo
  {pages} {649} (\bibinfo {year} {1954})}\BibitemShut {NoStop}%
\bibitem [{\citenamefont {Ospelkaus}\ \emph {et~al.}(2008)\citenamefont
  {Ospelkaus}, \citenamefont {Langer}, \citenamefont {Amini}, \citenamefont
  {Brown}, \citenamefont {Leibfried},\ and\ \citenamefont
  {Wineland}}]{ospelkaus_2008}%
  \BibitemOpen
  \bibfield  {author} {\bibinfo {author} {\bibfnamefont {C.}~\bibnamefont
  {Ospelkaus}}, \bibinfo {author} {\bibfnamefont {C.}~\bibnamefont {Langer}},
  \bibinfo {author} {\bibfnamefont {J.~M.}\ \bibnamefont {Amini}}, \bibinfo
  {author} {\bibfnamefont {K.~R.}\ \bibnamefont {Brown}}, \bibinfo {author}
  {\bibfnamefont {D.}~\bibnamefont {Leibfried}}, \ and\ \bibinfo {author}
  {\bibfnamefont {D.~J.}\ \bibnamefont {Wineland}},\ }\href@noop {} {\bibfield
  {journal} {\bibinfo  {journal} {Phys. Rev. Lett.}\ }\textbf {\bibinfo
  {volume} {101}},\ \bibinfo {pages} {090502} (\bibinfo {year}
  {2008})}\BibitemShut {NoStop}%
\bibitem [{\citenamefont {Ospelkaus}\ \emph {et~al.}(2011)\citenamefont
  {Ospelkaus}, \citenamefont {Warring}, \citenamefont {Colombe}, \citenamefont
  {Brown}, \citenamefont {Amini}, \citenamefont {Leibfried},\ and\
  \citenamefont {Wineland}}]{ospelkaus_2011}%
  \BibitemOpen
  \bibfield  {author} {\bibinfo {author} {\bibfnamefont {C.}~\bibnamefont
  {Ospelkaus}}, \bibinfo {author} {\bibfnamefont {U.}~\bibnamefont {Warring}},
  \bibinfo {author} {\bibfnamefont {Y.}~\bibnamefont {Colombe}}, \bibinfo
  {author} {\bibfnamefont {K.~R.}\ \bibnamefont {Brown}}, \bibinfo {author}
  {\bibfnamefont {J.~M.}\ \bibnamefont {Amini}}, \bibinfo {author}
  {\bibfnamefont {D.}~\bibnamefont {Leibfried}}, \ and\ \bibinfo {author}
  {\bibfnamefont {D.~J.}\ \bibnamefont {Wineland}},\ }\href@noop {} {\bibfield
  {journal} {\bibinfo  {journal} {Nature}\ }\textbf {\bibinfo {volume} {476}},\
  \bibinfo {pages} {181} (\bibinfo {year} {2011})}\BibitemShut {NoStop}%
\bibitem [{\citenamefont {Roos}(2008)}]{roos_2008}%
  \BibitemOpen
  \bibfield  {author} {\bibinfo {author} {\bibfnamefont {C.~F.}\ \bibnamefont
  {Roos}},\ }\href@noop {} {\bibfield  {journal} {\bibinfo  {journal} {New J.
  Phys.}\ }\textbf {\bibinfo {volume} {10}},\ \bibinfo {pages} {013002}
  (\bibinfo {year} {2008})}\BibitemShut {NoStop}%
\bibitem [{\citenamefont {Jonathan}\ \emph {et~al.}(2000)\citenamefont
  {Jonathan}, \citenamefont {Plenio},\ and\ \citenamefont
  {Knight}}]{jonathan_2000}%
  \BibitemOpen
  \bibfield  {author} {\bibinfo {author} {\bibfnamefont {D.}~\bibnamefont
  {Jonathan}}, \bibinfo {author} {\bibfnamefont {M.~B.}\ \bibnamefont
  {Plenio}}, \ and\ \bibinfo {author} {\bibfnamefont {P.~L.}\ \bibnamefont
  {Knight}},\ }\href {\doibase 10.1103/PhysRevA.62.042307} {\bibfield
  {journal} {\bibinfo  {journal} {Phys. Rev. A}\ }\textbf {\bibinfo {volume}
  {62}},\ \bibinfo {pages} {042307} (\bibinfo {year} {2000})}\BibitemShut
  {NoStop}%
\bibitem [{\citenamefont {Sutherland}\ \emph {et~al.}(2020)\citenamefont
  {Sutherland}, \citenamefont {Srinivas}, \citenamefont {Burd}, \citenamefont
  {Knaack}, \citenamefont {Wilson}, \citenamefont {Wineland}, \citenamefont
  {Leibfried}, \citenamefont {Allcock}, \citenamefont {Slichter},\ and\
  \citenamefont {Libby}}]{sutherland_2020}%
  \BibitemOpen
  \bibfield  {author} {\bibinfo {author} {\bibfnamefont {R.~T.}\ \bibnamefont
  {Sutherland}}, \bibinfo {author} {\bibfnamefont {R.}~\bibnamefont
  {Srinivas}}, \bibinfo {author} {\bibfnamefont {S.~C.}\ \bibnamefont {Burd}},
  \bibinfo {author} {\bibfnamefont {H.~M.}\ \bibnamefont {Knaack}}, \bibinfo
  {author} {\bibfnamefont {A.~C.}\ \bibnamefont {Wilson}}, \bibinfo {author}
  {\bibfnamefont {D.~J.}\ \bibnamefont {Wineland}}, \bibinfo {author}
  {\bibfnamefont {D.}~\bibnamefont {Leibfried}}, \bibinfo {author}
  {\bibfnamefont {D.~T.~C.}\ \bibnamefont {Allcock}}, \bibinfo {author}
  {\bibfnamefont {D.~H.}\ \bibnamefont {Slichter}}, \ and\ \bibinfo {author}
  {\bibfnamefont {S.~B.}\ \bibnamefont {Libby}},\ }\href {\doibase
  10.1103/PhysRevA.101.042334} {\bibfield  {journal} {\bibinfo  {journal}
  {Phys. Rev. A}\ }\textbf {\bibinfo {volume} {101}},\ \bibinfo {pages}
  {042334} (\bibinfo {year} {2020})}\BibitemShut {NoStop}%
\bibitem [{\citenamefont {Caves}(1981)}]{caves_1981}%
  \BibitemOpen
  \bibfield  {author} {\bibinfo {author} {\bibfnamefont {C.~M.}\ \bibnamefont
  {Caves}},\ }\href@noop {} {\bibfield  {journal} {\bibinfo  {journal} {Phys.
  Rev. D}\ }\textbf {\bibinfo {volume} {23}},\ \bibinfo {pages} {1693}
  (\bibinfo {year} {1981})}\BibitemShut {NoStop}%
\bibitem [{\citenamefont {Aasi}\ \emph {et~al.}(2013)\citenamefont {Aasi} \emph
  {et~al.}}]{aasi_2013}%
  \BibitemOpen
  \bibfield  {author} {\bibinfo {author} {\bibfnamefont {J.}~\bibnamefont
  {Aasi}} \emph {et~al.},\ }\href@noop {} {\bibfield  {journal} {\bibinfo
  {journal} {Nat. Photon.}\ }\textbf {\bibinfo {volume} {7}},\ \bibinfo {pages}
  {613} (\bibinfo {year} {2013})}\BibitemShut {NoStop}%
\bibitem [{\citenamefont {Ge}\ \emph {et~al.}(2019)\citenamefont {Ge},
  \citenamefont {Sawyer}, \citenamefont {Britton}, \citenamefont {Jacobs},
  \citenamefont {Bollinger},\ and\ \citenamefont {Foss-Feig}}]{ge_2019}%
  \BibitemOpen
  \bibfield  {author} {\bibinfo {author} {\bibfnamefont {W.}~\bibnamefont
  {Ge}}, \bibinfo {author} {\bibfnamefont {B.~C.}\ \bibnamefont {Sawyer}},
  \bibinfo {author} {\bibfnamefont {J.~W.}\ \bibnamefont {Britton}}, \bibinfo
  {author} {\bibfnamefont {K.}~\bibnamefont {Jacobs}}, \bibinfo {author}
  {\bibfnamefont {J.~J.}\ \bibnamefont {Bollinger}}, \ and\ \bibinfo {author}
  {\bibfnamefont {M.}~\bibnamefont {Foss-Feig}},\ }\href@noop {} {\bibfield
  {journal} {\bibinfo  {journal} {Phys. Rev. Lett.}\ }\textbf {\bibinfo
  {volume} {122}},\ \bibinfo {pages} {030501} (\bibinfo {year}
  {2019})}\BibitemShut {NoStop}%
\bibitem [{\citenamefont {Burd}\ \emph {et~al.}(2020)\citenamefont {Burd},
  \citenamefont {Srinivas}, \citenamefont {Knaack}, \citenamefont {Ge},
  \citenamefont {Wilson}, \citenamefont {Wineland}, \citenamefont {Leibfried},
  \citenamefont {Bollinger}, \citenamefont {Allcock},\ and\ \citenamefont
  {Slichter}}]{burd_2020}%
  \BibitemOpen
  \bibfield  {author} {\bibinfo {author} {\bibfnamefont {S.~C.}\ \bibnamefont
  {Burd}}, \bibinfo {author} {\bibfnamefont {R.}~\bibnamefont {Srinivas}},
  \bibinfo {author} {\bibfnamefont {H.~M.}\ \bibnamefont {Knaack}}, \bibinfo
  {author} {\bibfnamefont {W.}~\bibnamefont {Ge}}, \bibinfo {author}
  {\bibfnamefont {A.~C.}\ \bibnamefont {Wilson}}, \bibinfo {author}
  {\bibfnamefont {D.~J.}\ \bibnamefont {Wineland}}, \bibinfo {author}
  {\bibfnamefont {D.}~\bibnamefont {Leibfried}}, \bibinfo {author}
  {\bibfnamefont {J.~J.}\ \bibnamefont {Bollinger}}, \bibinfo {author}
  {\bibfnamefont {D.~T.~C.}\ \bibnamefont {Allcock}}, \ and\ \bibinfo {author}
  {\bibfnamefont {D.~H.}\ \bibnamefont {Slichter}},\ }\href@noop {} {\bibfield
  {journal} {\bibinfo  {journal} {arXiv preprint arXiv:2009.14342}\ } (\bibinfo
  {year} {2020})}\BibitemShut {NoStop}%
\bibitem [{\citenamefont {Wittemer}\ \emph {et~al.}(2019)\citenamefont
  {Wittemer}, \citenamefont {Hakelberg}, \citenamefont {Kiefer}, \citenamefont
  {Schr{\"o}der}, \citenamefont {Fey}, \citenamefont {Sch{\"u}tzhold},
  \citenamefont {Warring},\ and\ \citenamefont {Schaetz}}]{wittemer_2019}%
  \BibitemOpen
  \bibfield  {author} {\bibinfo {author} {\bibfnamefont {M.}~\bibnamefont
  {Wittemer}}, \bibinfo {author} {\bibfnamefont {F.}~\bibnamefont {Hakelberg}},
  \bibinfo {author} {\bibfnamefont {P.}~\bibnamefont {Kiefer}}, \bibinfo
  {author} {\bibfnamefont {J.-P.}\ \bibnamefont {Schr{\"o}der}}, \bibinfo
  {author} {\bibfnamefont {C.}~\bibnamefont {Fey}}, \bibinfo {author}
  {\bibfnamefont {R.}~\bibnamefont {Sch{\"u}tzhold}}, \bibinfo {author}
  {\bibfnamefont {U.}~\bibnamefont {Warring}}, \ and\ \bibinfo {author}
  {\bibfnamefont {T.}~\bibnamefont {Schaetz}},\ }\href@noop {} {\bibfield
  {journal} {\bibinfo  {journal} {Phys. Rev. Lett.}\ }\textbf {\bibinfo
  {volume} {123}},\ \bibinfo {pages} {180502} (\bibinfo {year}
  {2019})}\BibitemShut {NoStop}%
\bibitem [{\citenamefont {Dupays}\ and\ \citenamefont
  {Chenu}(2021)}]{dupays_2021}%
  \BibitemOpen
  \bibfield  {author} {\bibinfo {author} {\bibfnamefont {L.}~\bibnamefont
  {Dupays}}\ and\ \bibinfo {author} {\bibfnamefont {A.}~\bibnamefont {Chenu}},\
  }\href@noop {} {\bibfield  {journal} {\bibinfo  {journal} {Quantum}\ }\textbf
  {\bibinfo {volume} {5}},\ \bibinfo {pages} {449} (\bibinfo {year}
  {2021})}\BibitemShut {NoStop}%
\bibitem [{\citenamefont {Agarwal}(2012)}]{agarwal_2012}%
  \BibitemOpen
  \bibfield  {author} {\bibinfo {author} {\bibfnamefont {G.~S.}\ \bibnamefont
  {Agarwal}},\ }\href@noop {} {\emph {\bibinfo {title} {Quantum optics}}}\
  (\bibinfo  {publisher} {Cambridge University Press},\ \bibinfo {year}
  {2012})\BibitemShut {NoStop}%
\bibitem [{\citenamefont {Weedbrook}\ \emph {et~al.}(2012)\citenamefont
  {Weedbrook}, \citenamefont {Pirandola}, \citenamefont
  {Garc{\'\i}a-Patr{\'o}n}, \citenamefont {Cerf}, \citenamefont {Ralph},
  \citenamefont {Shapiro},\ and\ \citenamefont {Lloyd}}]{weedbrook_2012}%
  \BibitemOpen
  \bibfield  {author} {\bibinfo {author} {\bibfnamefont {C.}~\bibnamefont
  {Weedbrook}}, \bibinfo {author} {\bibfnamefont {S.}~\bibnamefont
  {Pirandola}}, \bibinfo {author} {\bibfnamefont {R.}~\bibnamefont
  {Garc{\'\i}a-Patr{\'o}n}}, \bibinfo {author} {\bibfnamefont {N.~J.}\
  \bibnamefont {Cerf}}, \bibinfo {author} {\bibfnamefont {T.~C.}\ \bibnamefont
  {Ralph}}, \bibinfo {author} {\bibfnamefont {J.~H.}\ \bibnamefont {Shapiro}},
  \ and\ \bibinfo {author} {\bibfnamefont {S.}~\bibnamefont {Lloyd}},\
  }\href@noop {} {\bibfield  {journal} {\bibinfo  {journal} {Rev. Mod. Phys.}\
  }\textbf {\bibinfo {volume} {84}},\ \bibinfo {pages} {621} (\bibinfo {year}
  {2012})}\BibitemShut {NoStop}%
\bibitem [{\citenamefont {Grosshans}\ \emph {et~al.}(2003)\citenamefont
  {Grosshans}, \citenamefont {Van~Assche}, \citenamefont {Wenger},
  \citenamefont {Brouri}, \citenamefont {Cerf},\ and\ \citenamefont
  {Grangier}}]{grosshans_2003}%
  \BibitemOpen
  \bibfield  {author} {\bibinfo {author} {\bibfnamefont {F.}~\bibnamefont
  {Grosshans}}, \bibinfo {author} {\bibfnamefont {G.}~\bibnamefont
  {Van~Assche}}, \bibinfo {author} {\bibfnamefont {J.}~\bibnamefont {Wenger}},
  \bibinfo {author} {\bibfnamefont {R.}~\bibnamefont {Brouri}}, \bibinfo
  {author} {\bibfnamefont {N.~J.}\ \bibnamefont {Cerf}}, \ and\ \bibinfo
  {author} {\bibfnamefont {P.}~\bibnamefont {Grangier}},\ }\href@noop {}
  {\bibfield  {journal} {\bibinfo  {journal} {Nature}\ }\textbf {\bibinfo
  {volume} {421}},\ \bibinfo {pages} {238} (\bibinfo {year}
  {2003})}\BibitemShut {NoStop}%
\bibitem [{\citenamefont {Furusawa}\ \emph {et~al.}(1998)\citenamefont
  {Furusawa}, \citenamefont {S{\o}rensen}, \citenamefont {Braunstein},
  \citenamefont {Fuchs}, \citenamefont {Kimble},\ and\ \citenamefont
  {Polzik}}]{furusawa_1998}%
  \BibitemOpen
  \bibfield  {author} {\bibinfo {author} {\bibfnamefont {A.}~\bibnamefont
  {Furusawa}}, \bibinfo {author} {\bibfnamefont {J.~L.}\ \bibnamefont
  {S{\o}rensen}}, \bibinfo {author} {\bibfnamefont {S.~L.}\ \bibnamefont
  {Braunstein}}, \bibinfo {author} {\bibfnamefont {C.~A.}\ \bibnamefont
  {Fuchs}}, \bibinfo {author} {\bibfnamefont {H.~J.}\ \bibnamefont {Kimble}}, \
  and\ \bibinfo {author} {\bibfnamefont {E.~S.}\ \bibnamefont {Polzik}},\
  }\href@noop {} {\bibfield  {journal} {\bibinfo  {journal} {Science}\ }\textbf
  {\bibinfo {volume} {282}},\ \bibinfo {pages} {706} (\bibinfo {year}
  {1998})}\BibitemShut {NoStop}%
\bibitem [{\citenamefont {Eichler}\ \emph {et~al.}(2011)\citenamefont
  {Eichler}, \citenamefont {Bozyigit}, \citenamefont {Lang}, \citenamefont
  {Baur}, \citenamefont {Steffen}, \citenamefont {Fink}, \citenamefont
  {Filipp},\ and\ \citenamefont {Wallraff}}]{eichler_2011}%
  \BibitemOpen
  \bibfield  {author} {\bibinfo {author} {\bibfnamefont {C.}~\bibnamefont
  {Eichler}}, \bibinfo {author} {\bibfnamefont {D.}~\bibnamefont {Bozyigit}},
  \bibinfo {author} {\bibfnamefont {C.}~\bibnamefont {Lang}}, \bibinfo {author}
  {\bibfnamefont {M.}~\bibnamefont {Baur}}, \bibinfo {author} {\bibfnamefont
  {L.}~\bibnamefont {Steffen}}, \bibinfo {author} {\bibfnamefont {J.~M.}\
  \bibnamefont {Fink}}, \bibinfo {author} {\bibfnamefont {S.}~\bibnamefont
  {Filipp}}, \ and\ \bibinfo {author} {\bibfnamefont {A.}~\bibnamefont
  {Wallraff}},\ }\href {\doibase 10.1103/PhysRevLett.107.113601} {\bibfield
  {journal} {\bibinfo  {journal} {Phys. Rev. Lett.}\ }\textbf {\bibinfo
  {volume} {107}},\ \bibinfo {pages} {113601} (\bibinfo {year}
  {2011})}\BibitemShut {NoStop}%
\bibitem [{\citenamefont {Gorman}\ \emph {et~al.}(2014)\citenamefont {Gorman},
  \citenamefont {Schindler}, \citenamefont {Selvarajan}, \citenamefont
  {Daniilidis},\ and\ \citenamefont {H{\"a}ffner}}]{gorman_2014}%
  \BibitemOpen
  \bibfield  {author} {\bibinfo {author} {\bibfnamefont {D.~J.}\ \bibnamefont
  {Gorman}}, \bibinfo {author} {\bibfnamefont {P.}~\bibnamefont {Schindler}},
  \bibinfo {author} {\bibfnamefont {S.}~\bibnamefont {Selvarajan}}, \bibinfo
  {author} {\bibfnamefont {N.}~\bibnamefont {Daniilidis}}, \ and\ \bibinfo
  {author} {\bibfnamefont {H.}~\bibnamefont {H{\"a}ffner}},\ }\href@noop {}
  {\bibfield  {journal} {\bibinfo  {journal} {Phys. Rev. A}\ }\textbf {\bibinfo
  {volume} {89}},\ \bibinfo {pages} {062332} (\bibinfo {year}
  {2014})}\BibitemShut {NoStop}%
\bibitem [{\citenamefont {Lau}(2014)}]{lau_2014}%
  \BibitemOpen
  \bibfield  {author} {\bibinfo {author} {\bibfnamefont {H.-K.}\ \bibnamefont
  {Lau}},\ }\href@noop {} {\bibfield  {journal} {\bibinfo  {journal} {Phys.
  Rev. A}\ }\textbf {\bibinfo {volume} {90}},\ \bibinfo {pages} {063401}
  (\bibinfo {year} {2014})}\BibitemShut {NoStop}%
\bibitem [{\citenamefont {Fisher}\ \emph {et~al.}(1984)\citenamefont {Fisher},
  \citenamefont {Nieto},\ and\ \citenamefont {Sandberg}}]{fisher_1984}%
  \BibitemOpen
  \bibfield  {author} {\bibinfo {author} {\bibfnamefont {R.~A.}\ \bibnamefont
  {Fisher}}, \bibinfo {author} {\bibfnamefont {M.~M.}\ \bibnamefont {Nieto}}, \
  and\ \bibinfo {author} {\bibfnamefont {V.~D.}\ \bibnamefont {Sandberg}},\
  }\href@noop {} {\bibfield  {journal} {\bibinfo  {journal} {Phys. Rev. D}\
  }\textbf {\bibinfo {volume} {29}},\ \bibinfo {pages} {1107} (\bibinfo {year}
  {1984})}\BibitemShut {NoStop}%
\bibitem [{\citenamefont {Hillery}\ \emph {et~al.}(1984)\citenamefont
  {Hillery}, \citenamefont {Zubairy},\ and\ \citenamefont
  {Wodkiewicz}}]{hillery_1984}%
  \BibitemOpen
  \bibfield  {author} {\bibinfo {author} {\bibfnamefont {M.}~\bibnamefont
  {Hillery}}, \bibinfo {author} {\bibfnamefont {M.}~\bibnamefont {Zubairy}}, \
  and\ \bibinfo {author} {\bibfnamefont {K.}~\bibnamefont {Wodkiewicz}},\
  }\href@noop {} {\bibfield  {journal} {\bibinfo  {journal} {Phys. Lett. A}\
  }\textbf {\bibinfo {volume} {103}},\ \bibinfo {pages} {259} (\bibinfo {year}
  {1984})}\BibitemShut {NoStop}%
\bibitem [{\citenamefont {Braunstein}\ and\ \citenamefont
  {McLachlan}(1987)}]{braunstein_1987}%
  \BibitemOpen
  \bibfield  {author} {\bibinfo {author} {\bibfnamefont {S.~L.}\ \bibnamefont
  {Braunstein}}\ and\ \bibinfo {author} {\bibfnamefont {R.~I.}\ \bibnamefont
  {McLachlan}},\ }\href@noop {} {\bibfield  {journal} {\bibinfo  {journal}
  {Phys. Rev. A}\ }\textbf {\bibinfo {volume} {35}},\ \bibinfo {pages} {1659}
  (\bibinfo {year} {1987})}\BibitemShut {NoStop}%
\bibitem [{\citenamefont {Hillery}(1990)}]{hillery_1990}%
  \BibitemOpen
  \bibfield  {author} {\bibinfo {author} {\bibfnamefont {M.}~\bibnamefont
  {Hillery}},\ }\href@noop {} {\bibfield  {journal} {\bibinfo  {journal} {Phys.
  Rev. A}\ }\textbf {\bibinfo {volume} {42}},\ \bibinfo {pages} {498} (\bibinfo
  {year} {1990})}\BibitemShut {NoStop}%
\bibitem [{\citenamefont {Chang}\ \emph {et~al.}(2020)\citenamefont {Chang},
  \citenamefont {Sab{\'\i}n}, \citenamefont {Forn-D{\'\i}az}, \citenamefont
  {Quijandr{\'\i}a}, \citenamefont {Vadiraj}, \citenamefont {Nsanzineza},
  \citenamefont {Johansson},\ and\ \citenamefont {Wilson}}]{chang_2020}%
  \BibitemOpen
  \bibfield  {author} {\bibinfo {author} {\bibfnamefont {C.~S.}\ \bibnamefont
  {Chang}}, \bibinfo {author} {\bibfnamefont {C.}~\bibnamefont {Sab{\'\i}n}},
  \bibinfo {author} {\bibfnamefont {P.}~\bibnamefont {Forn-D{\'\i}az}},
  \bibinfo {author} {\bibfnamefont {F.}~\bibnamefont {Quijandr{\'\i}a}},
  \bibinfo {author} {\bibfnamefont {A.}~\bibnamefont {Vadiraj}}, \bibinfo
  {author} {\bibfnamefont {I.}~\bibnamefont {Nsanzineza}}, \bibinfo {author}
  {\bibfnamefont {G.}~\bibnamefont {Johansson}}, \ and\ \bibinfo {author}
  {\bibfnamefont {C.}~\bibnamefont {Wilson}},\ }\href@noop {} {\bibfield
  {journal} {\bibinfo  {journal} {Phys. Rev. X}\ }\textbf {\bibinfo {volume}
  {10}},\ \bibinfo {pages} {011011} (\bibinfo {year} {2020})}\BibitemShut
  {NoStop}%
\bibitem [{\citenamefont {Gottesman}(1998)}]{gottesman_1998}%
  \BibitemOpen
  \bibfield  {author} {\bibinfo {author} {\bibfnamefont {D.}~\bibnamefont
  {Gottesman}},\ }\href@noop {} {\bibfield  {journal} {\bibinfo  {journal}
  {arXiv preprint quant-ph/9807006}\ } (\bibinfo {year} {1998})}\BibitemShut
  {NoStop}%
\bibitem [{\citenamefont {Bartlett}\ and\ \citenamefont
  {Sanders}(2002)}]{bartlett_2002}%
  \BibitemOpen
  \bibfield  {author} {\bibinfo {author} {\bibfnamefont {S.~D.}\ \bibnamefont
  {Bartlett}}\ and\ \bibinfo {author} {\bibfnamefont {B.~C.}\ \bibnamefont
  {Sanders}},\ }\href@noop {} {\bibfield  {journal} {\bibinfo  {journal} {Phys.
  Rev. A}\ }\textbf {\bibinfo {volume} {65}},\ \bibinfo {pages} {042304}
  (\bibinfo {year} {2002})}\BibitemShut {NoStop}%
\bibitem [{\citenamefont {Jost}\ \emph {et~al.}(2009)\citenamefont {Jost},
  \citenamefont {Home}, \citenamefont {Amini}, \citenamefont {Hanneke},
  \citenamefont {Ozeri}, \citenamefont {Langer}, \citenamefont {Bollinger},
  \citenamefont {Leibfried},\ and\ \citenamefont {Wineland}}]{jost_2009}%
  \BibitemOpen
  \bibfield  {author} {\bibinfo {author} {\bibfnamefont {J.~D.}\ \bibnamefont
  {Jost}}, \bibinfo {author} {\bibfnamefont {J.~P.}\ \bibnamefont {Home}},
  \bibinfo {author} {\bibfnamefont {J.~M.}\ \bibnamefont {Amini}}, \bibinfo
  {author} {\bibfnamefont {D.}~\bibnamefont {Hanneke}}, \bibinfo {author}
  {\bibfnamefont {R.}~\bibnamefont {Ozeri}}, \bibinfo {author} {\bibfnamefont
  {C.}~\bibnamefont {Langer}}, \bibinfo {author} {\bibfnamefont {J.~J.}\
  \bibnamefont {Bollinger}}, \bibinfo {author} {\bibfnamefont {D.}~\bibnamefont
  {Leibfried}}, \ and\ \bibinfo {author} {\bibfnamefont {D.~J.}\ \bibnamefont
  {Wineland}},\ }\href@noop {} {\bibfield  {journal} {\bibinfo  {journal}
  {Nature}\ }\textbf {\bibinfo {volume} {459}},\ \bibinfo {pages} {683}
  (\bibinfo {year} {2009})}\BibitemShut {NoStop}%
\end{thebibliography}%

\pagebreak
\widetext
\begin{center}
\textbf{\large Supplemental Materials}
\end{center}
\setcounter{equation}{0}
\setcounter{figure}{0}
\setcounter{table}{0}
\setcounter{page}{1}
\makeatletter
\renewcommand{\theequation}{S\arabic{equation}}
\renewcommand{\thefigure}{S\arabic{figure}}
\renewcommand{\bibnumfmt}[1]{[S#1]}
\renewcommand{\citenumfont}[1]{S#1}

\section*{Universal continuous variable quantum computation with linear spin-motion coupling}\label{sec:discrete_uni}

Reference~\cite{lau_2016} showed that a spin's non-linear interaction with a harmonic oscillator may be used to generate non-Gaussian motional states. We here show that, when combined with single qubit rotations on the Bloch sphere, the interaction used in geometric phase gates can achieve universal CVQC. This Hamiltonian takes the form:
\begin{eqnarray}\label{eq:lloyd_gen_lin_typo}
    \hat{H}(\phi,\alpha,j) = \hbar\Omega\hat{\sigma}_{\alpha}(\hat{a}^{\dagger}_{j}e^{i\phi} + \hat{a}_{j}e^{-i\phi}),
\end{eqnarray}
where $\Omega$ is the interaction Rabi frequency, $\phi$ is a phase, $\alpha\in\{x, y, z\}$, and $j$ corresponds to the motional mode. Following Ref.~\cite{lloyd_1999}, we can generate higher-order interactions from Eq.~(\ref{eq:lloyd_gen_lin_typo}) using the sequence:
\begin{eqnarray}\label{eq:lloyd_typo}
e^{-(i/\hbar)\hat{H}\delta t}e^{-(i/\hbar)\hat{H}^{\prime}\delta t}e^{(i/\hbar)\hat{H}\delta t}e^{(i/\hbar)\hat{H}^{\prime}\delta t} &=& e^{(\delta t/\hbar)^{2}[\hat{H}^{\prime},\hat{H}]} + \mathcal{O}([\delta t\Omega]^{3}). \nonumber \\
\end{eqnarray}
This sequence has been used in trapped ions to simulate a spin-1/2 particle in an external potential \cite{leibfried_2002}, but relied on interactions that were non-linear in $\hat{a}_{j}$ to achieve a universal continuous variable gate set. Here, we use interactions that are only linear in $\hat{a}_{j}$, relaxing the experimental requirements. 

Choosing $\hat{H}(\phi,\alpha,j)$ and $\hat{H}^{\prime} = \hat{H}(\phi^{\prime},\alpha^{\prime},j^{\prime})$, we can obtain a general $2^{nd}$ order gate interaction:
\begin{eqnarray}\label{eq:second_order_typo}
\hat{U}_{2g} &=& \exp\Big(-2i\delta t^{2}\Omega^{2}\varepsilon_{\alpha,\alpha^{\prime},\alpha^{\prime\prime}}\hat{\sigma}_{\alpha^{\prime\prime}}\Big\{i\delta_{jj^{\prime}}\sin(\phi-\phi^{\prime}) + (\hat{a}^{\dagger}_{j}e^{i\phi} +\hat{a}_{j}e^{-i\phi})(\hat{a}^{\dagger}_{j^{\prime}}e^{i\phi^{\prime}} + \hat{a}_{j^{\prime}}e^{-i\phi^{\prime}})\Big\}\Big) + \mathcal{O}([\delta t\Omega]^{3}),
\end{eqnarray}
where $\delta_{jj^{\prime}}$ is the Kronecker delta function, and $\varepsilon_{\alpha,\alpha^{\prime},\alpha^{\prime\prime}}$ is the Levi-Civita symbol. If we assume $\alpha\neq \alpha^{\prime}$, and apply a single-qubit rotation to transform the spin to a $\mp 1$ eigenstate of $\hat{\sigma}_{\alpha^{\prime\prime}}$, and ignore a global phase, the gate operator becomes:
\begin{eqnarray}
\hat{U}_{2g} &= \exp\Big(\pm 2i\delta t^{2}\Omega^{2}\Big\{&(\hat{a}^{\dagger}_{j}e^{i\phi} +\hat{a}_{j}e^{-i\phi})(\hat{a}^{\dagger}_{j^{\prime}}e^{i\phi^{\prime}} + \hat{a}_{j^{\prime}}e^{-i\phi^{\prime}})\Big\}\Big) + \mathcal{O}(\delta t^{3}).
\end{eqnarray}
In combination with coherent displacements , which can be generated by Eq.~(\ref{eq:lloyd_gen_lin_typo}) alone, this represents the Gaussian operations needed for universal CVQC \cite{weedbrook_2012}. Gaussian operations alone are not sufficient for universal quantum computation \cite{lloyd_1999}, however, and can be classically simulated efficiently if the initial state is Gaussian \cite{gottesman_1998,bartlett_2002}. To achieve higher-order motional couplings, we can apply Eq.~(\ref{eq:lloyd_typo}) again, substituting $\hat{U}_{2g}$ for $e^{\pm(\delta t/\hbar)^{2}\hat{H}^{\prime}}$, where the required $\hat{H}^{\prime}\rightarrow -\hat{H}^{\prime}$ may be achieved by applying $\pi$ pulses to the spin \cite{jost_2009}. This interaction will generate the degree-$3$ motional coupling needed to satisfy the Lloyd-Braunstein criterion for universal CVQC \cite{lloyd_1999}. For example, if we generate $\hat{U}_{2g}$ with $\hat{H}(0,x,j)$ and $\hat{H}^{\prime}(0,y,j)$ and, again, apply Eq.~(\ref{eq:lloyd_typo}) using $\hat{H} = \hat{H}(0,x,j)$ we get:
\begin{equation}
    \hat{U}_{3g} = \exp\Big(4i\delta t^{3}\Omega^{3}[\hat{a}^{\dagger}_{j}+\hat{a}_{j}]^{3}\hat{\sigma}_{y} \Big).
\end{equation}
If this operator acts on a $\pm 1$ eigenstate of $\hat{\sigma}_{y}$, we get:
\begin{equation}
    \hat{U}_{3g} \rightarrow \exp\Big(\pm 4i\delta t^{3}\Omega^{3}[\hat{a}^{\dagger}_{j}+\hat{a}_{j}]^{3}\Big).
\end{equation}
This time evolution operator corresponds to a cubic phase gate \cite{weedbrook_2012}, and, taken in combination with Gaussian operations, satisfies the Lloyd-Braunstein criterion. Note that the cubic phase gate was chosen for simplicity, and a similar prescription could be chosen to generate a broad set of degree-$3$ gates through different choices of $\phi$ and $j$ at each step. 

This section shows that our hybrid approach to universal CVQC has the same physical gate requirements as discrete quantum computing. There are, however, many situations where a digitized series of pulses is not desirable, for example if Gaussian interactions are needed for reasons other than CVQC \cite{heinzen_1990,burd_2019,burd_2020}. Therefore, in the main manuscript we discuss a technique for generating higher-order spin-motion coupling that uses continuously applied fields. 

\end{document}